\documentclass[aps,prd,nofootinbib,showpacs,floatfix,twocolumn]{revtex4}  
\usepackage{bm}
\usepackage{latexsym}
\usepackage{dcolumn}
\usepackage{amsfonts,amssymb,amsmath}
\usepackage{graphicx,epsfig}
\usepackage{psfrag}
\usepackage{subfigure}
\usepackage{feynmf}
\usepackage{rotating}
\usepackage{hyperref}
\usepackage{enumitem}
\usepackage{comment}
\usepackage{color}
\usepackage{makecell}
\DeclareMathOperator\erf{erf}
\usepackage[normalem]{ulem}
\usepackage[export]{adjustbox}

\hypersetup{
    unicode=false,          
    pdftoolbar=true,        
    pdfmenubar=true,        
    pdffitwindow=false,     
    pdfstartview={FitH},    
    pdftitle={My title},    
    pdfauthor={Author},     
    pdfsubject={Subject},   
    pdfcreator={Creator},   
    pdfproducer={Producer}, 
    pdfkeywords={keyword1} {key2} {key3}, 
    pdfnewwindow=true,      
    colorlinks=true,       
    linkcolor=blue,          
    citecolor=cyan,        
    filecolor=magenta,      
    urlcolor=blue,           
    linktocpage=true
}

\def\beq{\begin{equation}}
\def\eeq{\end{equation}}
\def\br{\begin{eqnarray}}
\def\er{\end{eqnarray}}
\def\benu{\begin{enumerate}}
\def\efnu{\end{enumerate}}

\def\l{\left}
\def\r{\right}

\usepackage{soul}

\begin{document}
\title{The asymmetry of dawn: evidence for asymmetric reionization histories from a joint analysis of cosmic microwave background and astrophysical data}
\author{Daniela Paoletti} \email{daniela.paoletti@inaf.it}
\affiliation{INAF OAS Bologna, Osservatorio di Astrofisica e Scienza dello Spazio di Bologna, via Gobetti 101, I-40129 Bologna, Italy  \\INFN, Sezione di Bologna, via Irnerio 46, 40126 Bologna, Italy}
\author{Dhiraj Kumar Hazra} \email{dhiraj@imsc.res.in}
\affiliation{The  Institute  of  Mathematical  Sciences,  HBNI,  CIT  Campus, Chennai  600113,  India\\ INAF OAS Bologna, Osservatorio di Astrofisica e Scienza dello Spazio di Bologna, via Gobetti 101, I-40129 Bologna, Italy}
\author{Fabio Finelli} \email{fabio.finelli@inaf.it}
\affiliation{INAF OAS Bologna, Osservatorio di Astrofisica e Scienza dello Spazio di Bologna, via Gobetti 101, I-40129 Bologna, Italy\\INFN, Sezione di Bologna, via Irnerio 46, 40126 Bologna, Italy}
\author{George F. Smoot}\email{gfsmoot@lbl.gov}
\affiliation{IAS TT \& WF Chao Foundation Professor,emertius, IAS, Hong Kong University of Science and Technology, Clear Water Bay, Kowloon, 999077 Hong Kong, China. \\
Paris Centre for Cosmological Physics, Universit\'{e} de Paris, emeritus, CNRS,  Astroparticule et Cosmologie, F-75013 Paris, France A, 10 rue Alice Domon et Leonie Duquet,75205 Paris CEDEX 13, France. \\
Donostia International Physics Center (DIPC), 20018 Donostia, The Basque Country, Spain. \\
Energetic Cosmos Laboratory, Nazarbayev University, Astana, Kazakhstan}
\date{\today}

\begin{abstract}
We show that by jointly fitting cosmic microwave background (CMB) and astrophysical data - a compilation of UV luminosity data from the Hubble Frontier Field and neutral hydrogen data from distant sources-, we can infer on the shape of the evolution of the ionized hydrogen fraction with redshift in addition to constraining the average optical depth $\tau$.
For this purpose, we introduce here a novel extended model 
that includes hydrogen ionization histories which are monotonic with redshift, but allow for an asymmetry as indicated from our previous works on a free reconstruction of reionization. By using our baseline data combination, we obtain $\tau=0.0542^{+0.0017}_{-0.0028}$, consistent with our previous works and tighter than the one inferred by Planck 2018 data because of the combination of CMB with astrophysical data. We find that the symmetric hypothesis within our parametrization is disfavoured at 4 $\sigma$.
We test our findings by using alternative likelihoods for CMB polarization at low multipoles, i.e. based on the 2020 reprocessing of Planck HFI data or on the joint analysis of WMAP and Planck LFI data, obtaining consistent results that disfavour the symmetric hypothesis of the reionization history at high statistical significant level.
These results will be further tested by more precise astrophysical data such as from JWST and Euclid deep fields.   
\end{abstract}
\maketitle
\section{Introduction}

Reionization represents one of the main phase transitions the Universe underwent in its entire history. With the first stars, dawn falls, the Universe gets ionized again long after the cosmological recombination of hydrogen and the dark ages end. We are, however, yet to understand the exact dynamics and origin of reionization, were they driven by the elusive POPIII stars, active galactic nuclei or through other mechanisms~\citep{Kulkarni:2018erh,Dayal2018,Fan2006:annurev,10.23943/princeton/9780691144917.001.0001,Barkana:2000fd}.

In the minimal assumptions for cosmological studies, the free electron fraction during reionization is usually modelled with a simplified, sharp transition in the form of an hyperbolic tangent with a fixed growth of the ionization fraction. In this way,reionization is simply encoded by one parameter, either the average optical depth $\tau$ or the effective redshift $z_\mathrm{re}$ at which the ionization fraction changes.With the recent progress in measuring CMB polarization on large angular scales~\citep{Planck2015:like,Planck:2016kqe,Pagano:2019tci,Delouis:2019bub,Natale:WMAPLFI,Planck2018:like}, $\tau$ remains the parameter with the largest uncertainties among the six ones which determine the concordance cosmological $\Lambda$CDM model as seen by Planck~\citep{Planck2018:param}, although its value has been reconciled with astrophysical measurements of the ionized medium~\citep{Robertson2015}.

The consistency of the inference of the average optical depth from CMB and astrophysical measurements under minimal assumptions has fuelled a renewed interest in the study of reionization beyond the average optical depth $\tau$~\citep{HS17,Paoletti:2020ndu} and in constraining the physics of the reionization process~\citep{PRL,PRD}. This is particularly important, since reionization can also be a source of confusion in extensions of $\Lambda$CDM, as for the neutrino mass~\citep{Archidiacono:2010wp,PRD}, annihilating dark matter~\citep{Paoletti:2020ndu} and most notably for certain inflationary motivated deviation from a power law for the primordial spectrum of curvature perturbations~\citep{HPBFSSS}.

As alternative to direct modelling of the free electron fraction \citep{HS17,Millea:2018bko,Paoletti:2020ndu}, reionization history can be obtained as a solution to the ionizing equations. In this way, on top of CMB anisotropy data, in particular the E-mode polarization, that constrains the integrated optical depth and to a certain extent also the shape and duration of the reionization, the following measurements can also be used:

\begin{enumerate}

\item UV luminosity data from Hubble Frontier Field~\cite{HFFsite,Coe2015,Lotz2017} that constrains the source part in the ionizing equation.

\item Neutral hydrogen fraction measurement from quasars, Gamma Ray Bursts (GRBs) and galaxies~\cite{Fan2006:annurev,Fan2006,Mortlock2011,Bolton2011,McGreer:2014qwa, Totani:2005ng,McQuinn:2007gm, Schroeder2013,Greig:2016vpu,Davies:2018pdw,McQuinn:2007dy,Ouchi2010,Ono2012,Caruana2013,Schenker2014,Tilvi2014,Pentericci2014,Sobacchi2015,Mason:2017eqr,Mason:2019ixe} that constrains the neutral/ionized hydrogen fraction obtained upon integrating the ionization equation.
\end{enumerate}

Several reionization models were therefore constrained using combinations of different data types~\cite{Robertson2010:Nature,Mitra:2011uv,Bouwens2015,Gorce2017,Ishigaki2018}. In general, astrophysical data carry an information which leads to resulting constraints on the optical depth which are tighter than those achievable by current CMB data alone. In~\cite{PRL} we introduced a free-form modelling of reionization and even within this conservative approach, we demonstrated that the reionization history can be constrained better than with a cosmic variance limited CMB observation alone, with the optical depth constrained up to 2\% accuracy.

The reconstructed histories of ionization found in~\cite{PRL} and other similar analyses do not show any preference for a symmetric model of ionization. Indeed, a symmetrical evolution of the free electron fraction does not arise from theoretical priors and it is a purely phenomenological choice for CMB parameter estimation. Since different reionization histories can lead to different constraints on the optical depth
and bias the estimation of other cosmological parameters, it is important to understand how current astrophysical and cosmological data have changed our view on the shapes of reionization history.

Motivated by our previous results, we study how distinguishing between symmetric and asymmetric models of reionization history with current CMB and astrophysics data. We introduce an extended model that allows different monotonic shapes of ionization evolution (ionization/free electron fraction) were the symmetric history can be obtained as the limit where an asymmetry parameter vanishes. With different data combinations we constrain this asymmetry parameter alongside the cosmological and other related parameters on reionization. As we mentioned before, it is important to have joint analysis with combined datasets in order to obtain stringent constraints. Since our starting point of this analysis is modelling the free electron fraction, in order to use the UV luminosity data, we reverse engineer the luminosity density from the ionization fraction. We test our results with different available large scale CMB polarization data and with different UV magnitude cuts in luminosity function. We also compare the Tanh model in particular against the extended model in the scale of Bayesian evidence.

The paper is organized as follows. In~\autoref{sec:Models} we discuss the extended model and its comparison with the baseline hyperbolic tangent model. In~\autoref{sec:Data} we discuss different dataset combinations and analysis details.~\autoref{sec:results-I} and~\autoref{sec:results-II} discuss the results of our analysis in two parts wherein, in the first part, we discuss the preference for the shapes of the ionization history from the data combinations and in the second part we do model comparison between our extended model and Tanh model. In~\autoref{sec:summary}, we summarize.

\begin{figure*}
\includegraphics[width=0.5\textwidth]{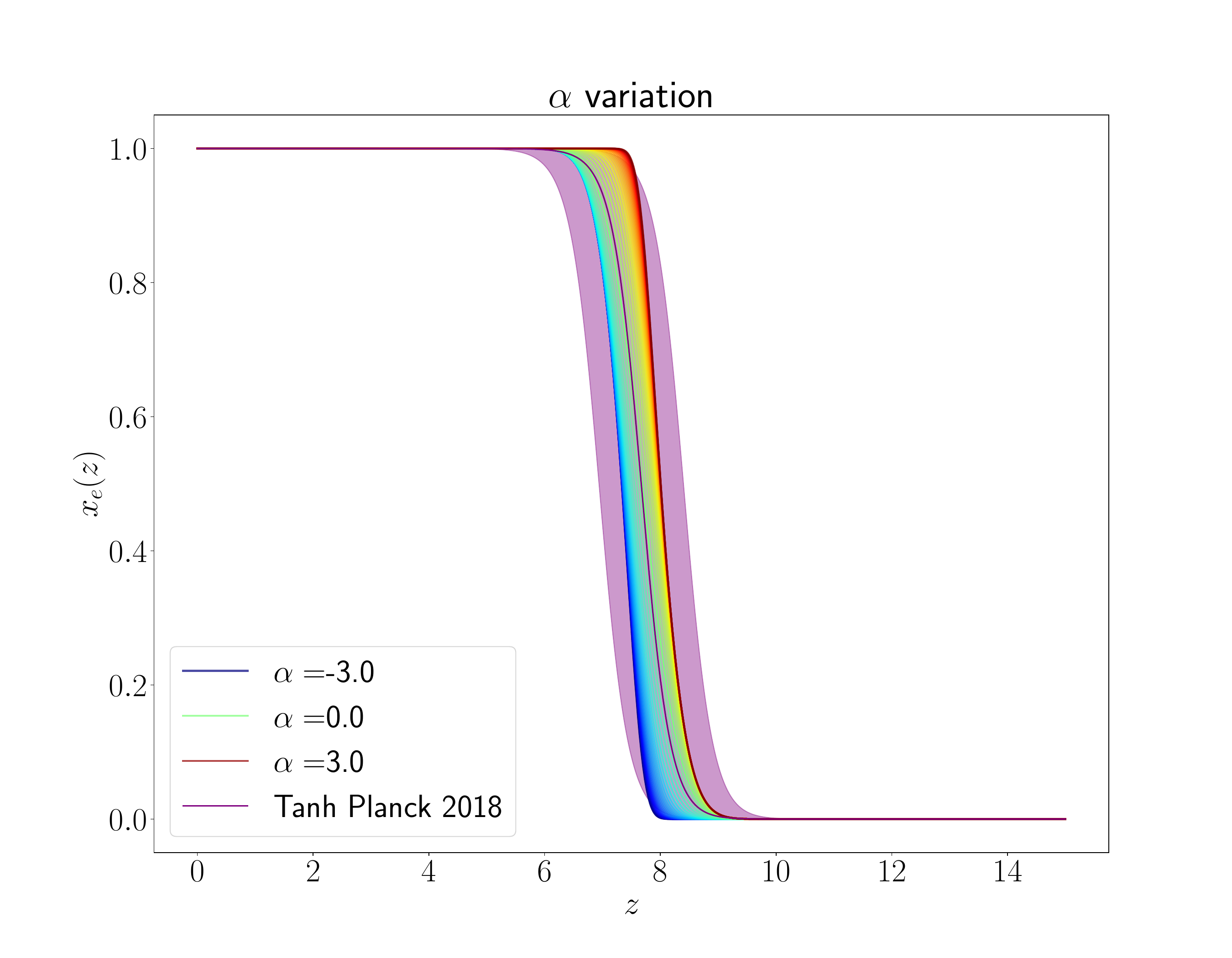}\includegraphics[width=0.5\textwidth]{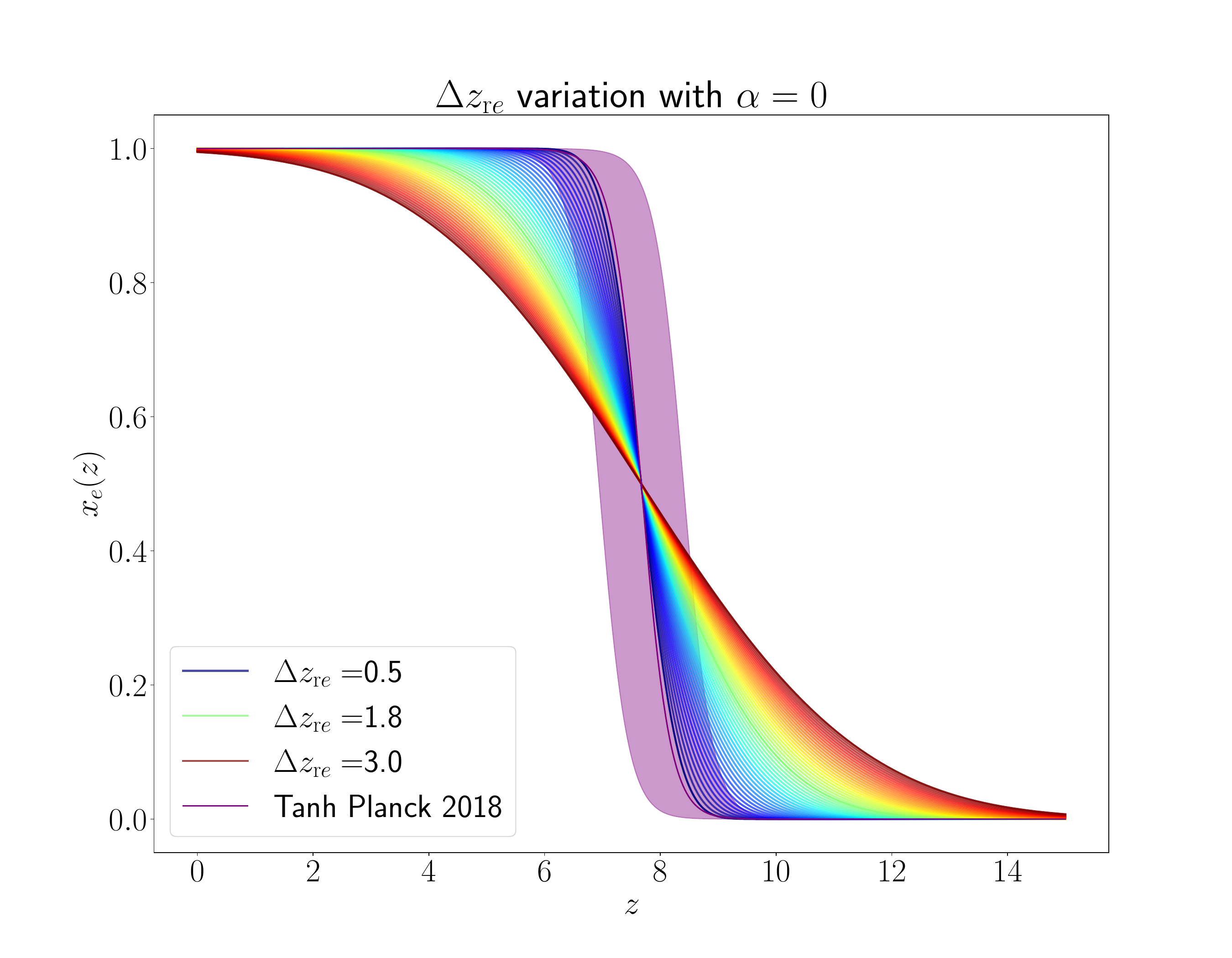}\\
\includegraphics[width=0.5\textwidth]{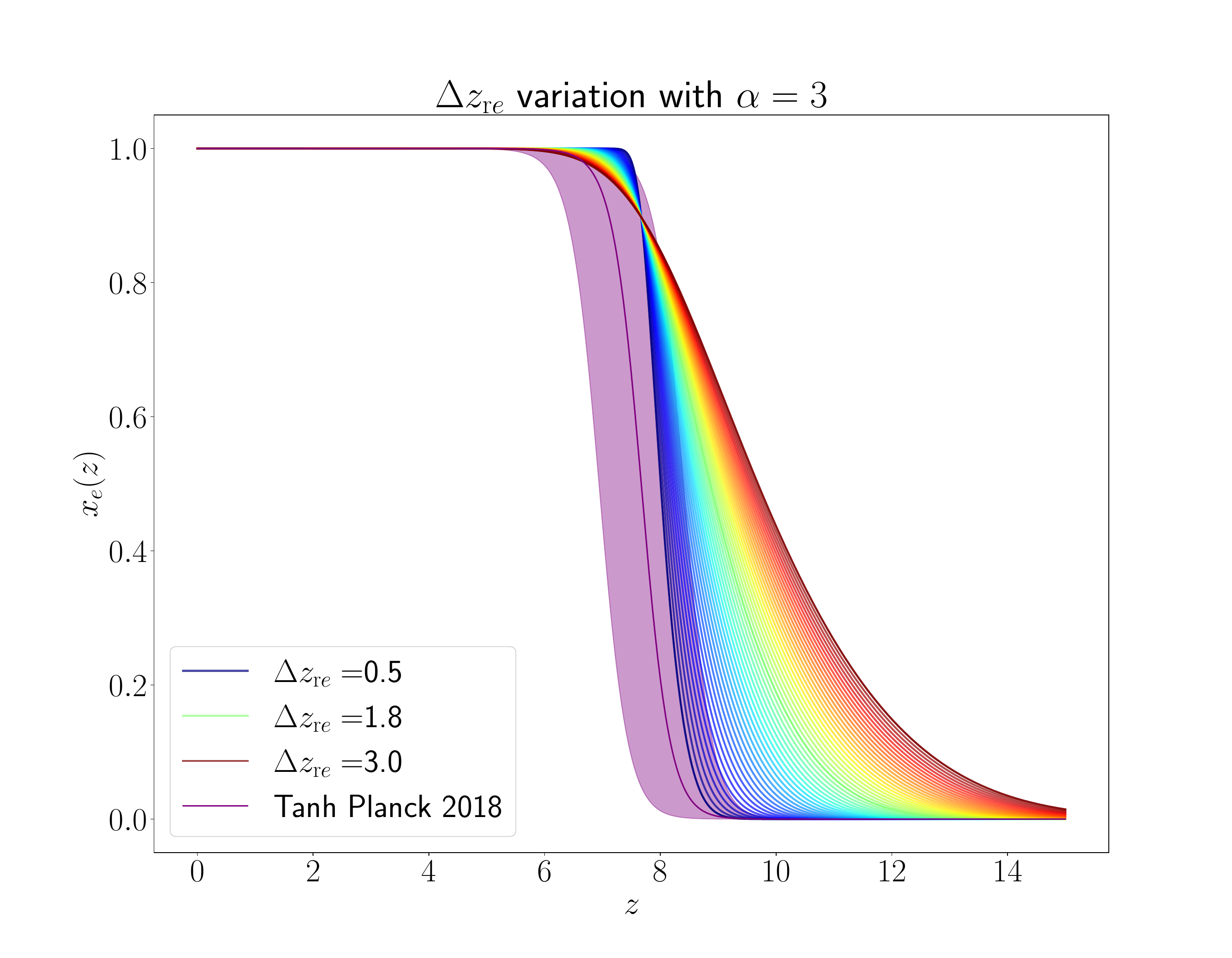}\includegraphics[width=0.5\textwidth]{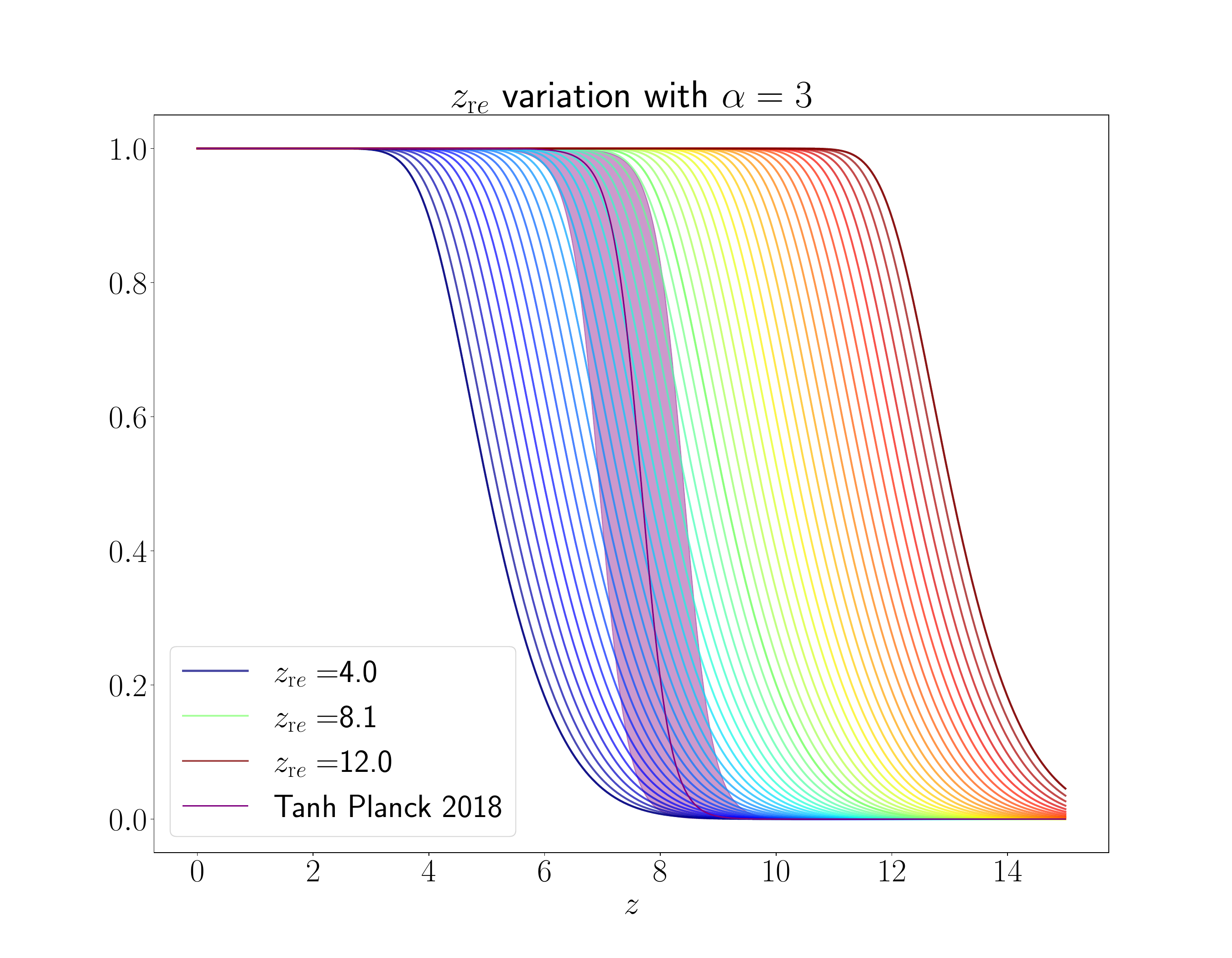}\\\includegraphics[width=0.5\textwidth]{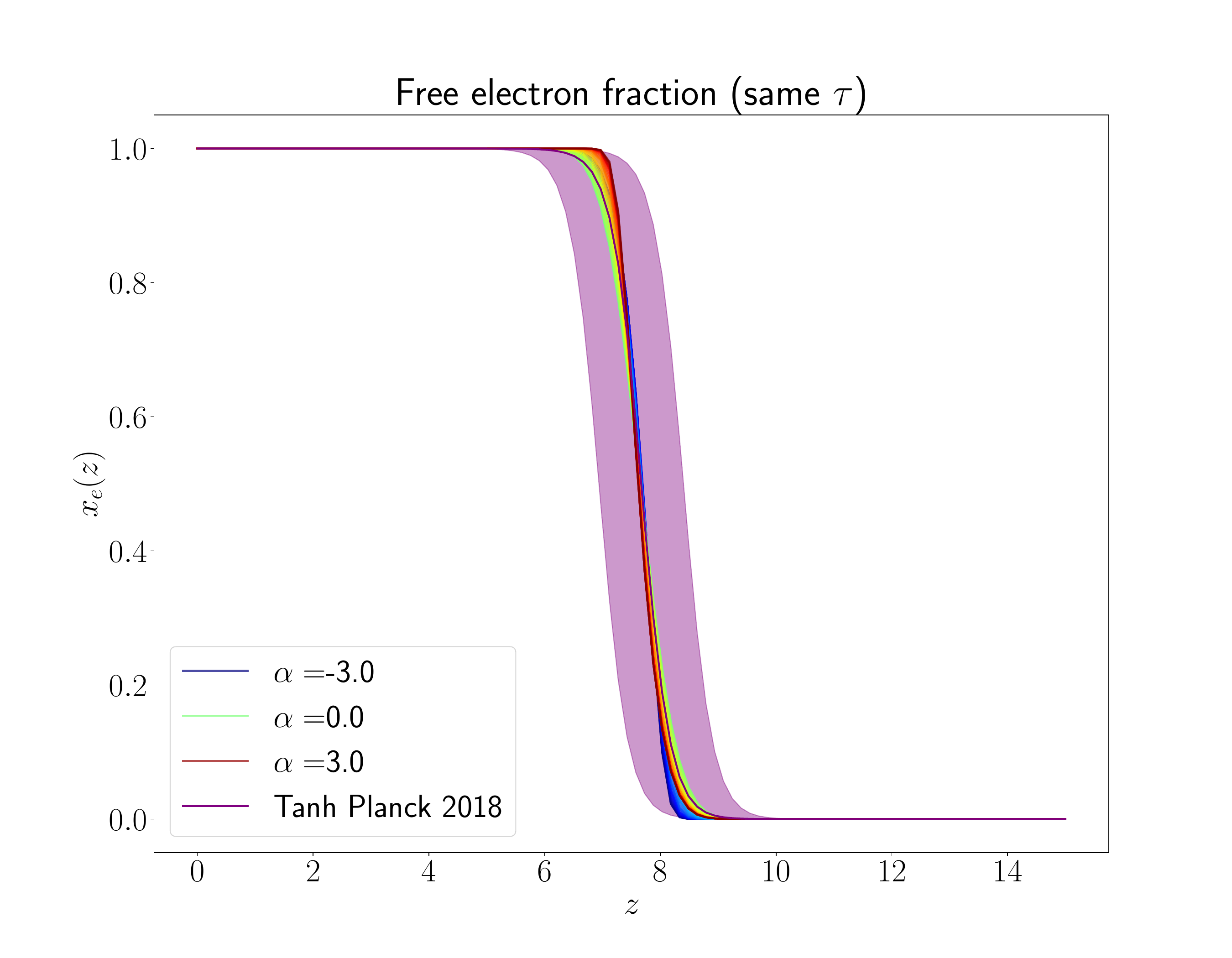}\includegraphics[width=0.5\textwidth]{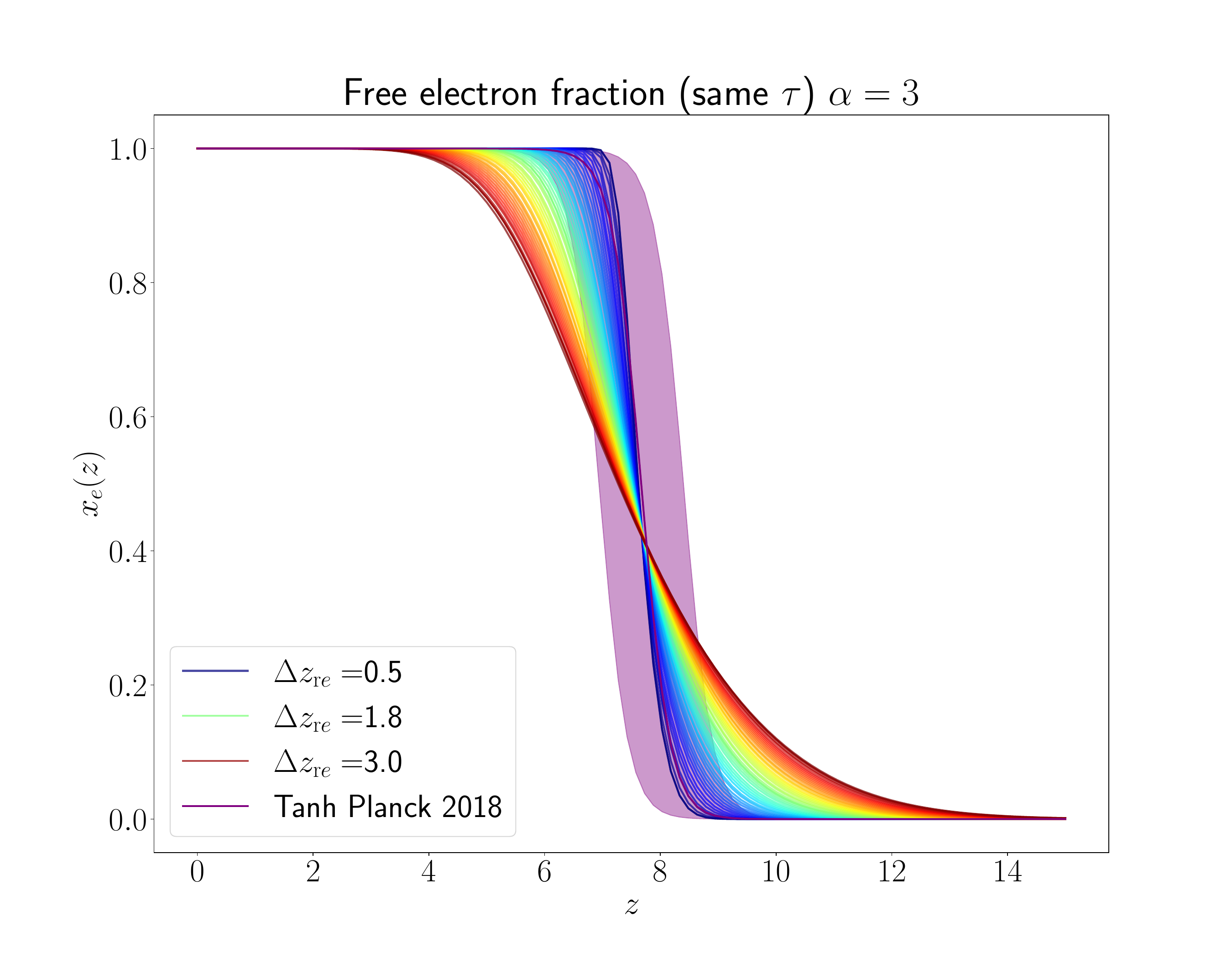}
\caption{Reionization histories plotted for the variation of the reionization parameters. Best fit $x^{\rm Tanh}_e(z)$ (Planck baseline)  from~\autoref{eq:Tanh} is plotted in purple in all plots. A resembling history from symmetric history as expressed in~\autoref{eq:Symmetric} is plotted. In the first four plots (in order of reading English books), keeping all other parameters fixed, we vary each parameter of the extended model (\autoref{eq:extended}) and plot the histories. In the top left plot we demonstrate the effect of varying $\alpha$ where departure from symmetry can be noticed. In the next three plots we vary reionization redshift and width. Top right plot shows the variation of reionization redshift. Bottom plots show the variation of the histories by varying $\alpha$ and $\Delta z_{\rm re}$ keeping the same optical depth. These two plots showcase certain cases that may be indistinguishable by Planck CMB data but can be differentiated by other observations.}\label{Fig:CDFDep}
\end{figure*}

\section{The model and methods}~\label{sec:Models}
\subsection{Extended model of ionizing history}

The straightforward method to describe reionization is to model the evolution with time or redshift of the average ionization fraction of the Universe with some function.
The standard model reionization in cosmology  assumes a transition from neutral to fully ionized in the form of an hyperbolic tangent in the ionization fraction:
\begin{equation}
        x^{\rm Tanh}_e(z)=\frac{1}{2}\left[1+\tanh\left(\frac{y(z_{\rm re})-y}{\Delta_y}\right)\right],~\label{eq:Tanh}
\end{equation}
where $y(z)=(1+z)^{3/2}$ and $\Delta_y = 3/2 \, \sqrt{1+z_{\rm re}} \, {\Delta}z_{\rm re} $. By definition this assumption imposes a symmetric nature of the reionization history.

We plan to address two related questions in this work,
\begin{enumerate}
    \item Do cosmological data from different sources and redshifts support a reionization history symmetric {\it w.r.t.} the redshift? 
    \item Is the widely used Tanh model, as provided in~\autoref{eq:Tanh} consistent with different cosmological data combinations?  
\end{enumerate}

In order to address the first question, we make use of an extended model where both symmetric and asymmetric models are nested within and a single parameter describes the asymmetry. Through this parameter  we can test the viability of a symmetric model. The second question can be addressed by model comparison.

\paragraph{Extended model:} Evolution of electron fraction in our model is described by a cumulative distribution function ($\Phi_{\rm Skewed}(z)$) corresponding to a skew normal distribution. $\Phi_{\rm Skewed}(z)$ is expressed with $\Phi_{\rm Normal}(z)$ as,
\begin{align}~\label{eq:SACDF}
    \Phi_{\rm Normal}(z)=\frac{1}{2}\left[1+\erf\left(\frac{z-z_{\rm re}}{\sqrt{2}\,{\Delta}z_{\rm re}}\right)\right]\\
    \Phi_{\rm Skewed}(z)=\Phi_{\rm Normal}(z)-2 T\left(\frac{z-z_{\rm re}}{{\Delta}z_{\rm re}},\alpha\right)
\end{align}
where T is the Owen T function.
The ionization fraction evolution that includes both symmetric and asymmetric evolution can be expressed in the extended model as, 
\begin{equation}~\label{eq:extended}
      x^{\rm Skewed}_e(z;z_{\rm re},{\Delta}z_{\rm re},\alpha)=1-\Phi_{\rm Skewed}(z;z_{\rm re},{\Delta}z_{\rm re},\alpha)  
\end{equation}
 Note that the distribution function depends on the reionization parameters and therefore in this equation we retained them as $\Phi_{\rm Skewed}(z)=\Phi_{\rm Skewed}(z;z_{\rm re},{\Delta}z_{\rm re},\alpha)$. Here $z_{\rm re},{\Delta}z_{\rm re}$ are the mean and standard deviation of the normal distribution that can be related to reionization redshift and width and $\alpha$ indicates the skewness or asymmetry in the distribution. For $\alpha=0$ the reionization history remains symmetric in redshift as, 
 \begin{equation}~\label{eq:Symmetric}
      x^{\rm Normal}_e(z;z_{\rm re},{\Delta}z_{\rm re})=1-\Phi_{\rm Normal}(z;z_{\rm re},{\Delta}z_{\rm re})  
\end{equation}

Note that this function does not exactly reproduce~\autoref{eq:Tanh}. However, by the variation of $z_{\rm re},{\Delta}z_{\rm re}$, it can be made to resemble $x^{\rm Tanh}_e(z)$. 
\paragraph{Parameter variation:}Since our analysis focuses on differentiating between the symmetric and asymmetric shapes of ionization history, in order to marginalize over the width of reionization, we also allow $\Delta z_{\rm re}$ to vary. Instead of sampling the reionization redshift ($z_{\rm re}$) which is normally associated to a symmetric history, we sample the optical depth to reionization, and $z_{\rm re}$ is obtained as a solution that satisfies the optical depth in that sample given the considered ionization history shape. In~\autoref{Fig:CDFDep}, we show the variation in the shape of ionization histories of our model, due to changes in the reionization parameters, the impact of the asymmetry parameter, the duration of reionization and the reionization redshift. 
We also show how with our extended model the same value of the optical depth can underlie different reionization histories stressing the importance of using different datasets to distinguish among them. In all the figures we show as a reference the Tanh model. 

The asymmetry parameter can be varied to have two branches of solutions around the symmetric history, namely, positive branch (corresponding to $\alpha>0$) and negative branch ($\alpha<0$).
Based on the hypothesis done on the possible sources of reionization and the results of our previous works on reconstructing the reionization history from its sources and CMB~\citep{PRL,PRD}, we do expect to have a history with a shape more compatible with the positive branch: a shallow transition from fully neutral to ionized and a sharper completion towards the end. On this basis, we will favour the positive branch for the choice of our priors. We have however tested that adding the negative branch prior provides similar constraints in the joint data combinations, Results are presented in \autoref{app}. 

\subsection{Reverse engineering the source from history: UV luminosity density}

In our earlier works on reconstructing the reionization history~\cite{PRL,PRD}, we started from the sources, integrating the free-form ionization and recombination rates to derive the reionization history. Here, since we start from the history, we need to reverse engineer the sources. 
Assuming the standard ionization equation, 
\begin{equation}~\label{eq:ionization}
\frac{dQ_{\rm H{II}}}{dt}=\frac{\dot{n}_{\rm ion}}{\langle n_H\rangle}-\frac{Q_{\rm H{II}}}{t_{\rm rec}}
\end{equation}
we trace back $\dot{n}_{\rm ion}$, given a form of $Q_{\rm H{II}}$. Note that in such treatments where the $Q_{\rm H{II}}$ has a known shape, its derivative $dQ_{\rm H{II}}/{dt}$ can be calculated analytically; in particular, for our extended model this derivative can be expressed simply in terms of a skewed normal. The recombination time is defined through the clumping factor $C_{\rm HII}$, recombination coefficient $\alpha_B(T)$, average density of hydrogen atom $\langle n_{\rm H}\rangle$, and hydrogen ($X_p$) and helium abundances ($Y_p$) and is expressed as \beq~\label{eq:recombination}
t_{\rm rec}(z)=\left[C_{\rm HII}\alpha_B(T)(1+\frac{Y_p}{4X_p})\langle n_{\rm H}\rangle (1+z)^3\right]^{-1}
\eeq
Here, $T$ refers to the temperature of the intergalactic medium and we we keep $T=20000K$ throughout our analysis.

Following the simulations in~\cite{Shull2012}, we use the following clumping factor expression,
\beq~\label{eq:clumping}
C_{\rm HII}(z)=2.9\l[\frac{1+z}{6}\r]^{-1.1}
\eeq

Using these expressions, for a given $\dot{n}_{\rm ion}$, obtained from the electron fraction model (equivalently the volume filling factor $Q_{\rm H{II}}$), UV luminosity density can be expressed as, 
\begin{equation}
\rho_{\rm UV}(z)=\frac{\dot{n}_{\rm ion}}{\langle f_{\rm esc}\dot{\xi}_{\rm reion}\rangle}\,.
\end{equation}

$\langle f_{\rm esc}\dot{\xi}_{\rm reion}\rangle$ represents the magnitude averaged value of the product of escape fraction $f_{\rm esc}$ and emission rate of photons. Note that CMB and neutral hydrogen fraction observations from quasars can only constrain the production rate of the ionizing photons $\dot{n}_{\rm ion}$ and they are blind towards individual terms within, since these observations probe only $Q_{\rm H{II}}(z)$. However, when luminosity function observations are used, they directly constrain $\rho_{\rm UV}$ which, in turn, helps us to constrain $\langle f_{\rm esc}\dot{\xi}_{\rm reion}\rangle$ when CMB and/or neutral hydrogen fraction data are used jointly. In our analysis we allow $\langle f_{\rm esc}\dot{\xi}_{\rm reion}\rangle$ 
to vary in combined analysis, where we allow this function to change over redshift monotonically. The amplitude is varied through $\dot{\xi}_{\rm reion}$ with a flat prior and the tilt is varied through the escape fraction with, 
\begin{equation}
    f_{\rm esc}(z)={\rm max}\l[\left(\frac{1+z}{15}\right)^\beta,1\r]
\end{equation}
using a flat prior on the slope $\beta$.

\section{DATA, PRIORS and SAMPLING}~\label{sec:Data}
The signatures of reionization are both {\it cosmological}, with a modification of the CMB anisotropy pattern in temperature and polarization, and {\it astrophysical} where it is possible to investigate reionization directly, with the measurements of the ionization fraction from distant objects, and indirectly, through the characterization of its sources with UV data.

Concerning the CMB, it is sensitive to reionization thanks to the Thomson scattering between photons and the newly free electrons during reionization. 

This effect damps the overall amplitude of the temperature anisotropy spectrum and it creates a bump in the $E$-mode polarization signal at large angular scales. This reionization bump, with its amplitude and shape, is sensitive to the optical depth and the duration of reionization history. However, being limited by cosmic variance, CMB can not probe the shape of the reionization histories, unless the reionization starts sufficiently early. However, such early onsets are disfavoured by the latest Planck data on polarization as we have shown in a model independent analysis~\cite{PRL}. 

In order to differentiate between the shapes, we complement CMB data with both direct ionization fraction measurements from sources as active galactic nuclei (AGNs) and GRBs and indirect information on the reionization sources coming from UV luminosity density using the same datasets~\cite{Ishigaki2018} as used in~\citep{PRL,PRD}.

\paragraph{CMB data:} As CMB data, we use Planck 2018 baseline which includes temperature, polarization and lensing likelihoods~\citep{Planck2018:like, Planck2018:lensing}. At $\ell \leq 30$, we use the temperature likelihood based on Gibbs sampling of the component separated map from {\texttt{commander}} and the $E$-mode polarization SIMALL likelihood. At high multipoles we use {\texttt{plik}} TTTEEE likelihood based on cross spectra. In the Planck baseline we also include the weak lensing likelihood in the conservative range $8 \le \ell \le 400$~\citep{Planck2018:lensing}. We also use alternative likelihoods at low-multipoles: SROLL2 \citep{Delouis:2019bub} and the joint WMAP-Planck LFI likelihood \citep{Natale:WMAPLFI}.

\paragraph{Neutral/ionization fraction data:} For the direct measurements of the ionization state around discrete sources as AGNs and GRBs the data spans a redshift range $z=6-8$. The different points used in our data combination consist of quasars and a GRB and can be found in ~\cite{McGreer:2014qwa,Totani:2005ng,McQuinn:2007gm,Schroeder2013,Greig:2016vpu,Davies:2018pdw,McQuinn:2007dy,Ouchi2010,Schenker2014,Tilvi2014,Mason:2017eqr,Mason:2019ixe}.

\paragraph{UV luminosity density:}For the UV luminosity density data we consider the  the Hubble Frontier Field~\cite{HFFsite,Lotz2017} observation of six galaxy clusters spanning redshift 6 to 11~\cite{Bouwens2014,Ishigaki2018}. Following our earlier work in~\citep{PRD}, we consider both a conservative cut of $-17$ for the magnitude of integration of the UV luminosity function and a more aggressive cut at $-15$ to check the impact of the different shape including the low magnitude tail of the luminosity function. It is important to highlight that the source function $\dot{n}_{\rm ion}(z)$ and therefore $\rho_{\rm UV}(z)$ can only be traced back during the reionization process, once $Q_{\rm H{II}}$ reaches unity we can no longer obtain $\dot{n}_{\rm ion}(z)$. Therefore, we do not use the UV luminosity data below $z=6.5$.

We consider different combinations of the data as described in~\autoref{tab:Datasets} to distinguish between the shapes of reionization history. Combinations using P2018, UV17 and QHII from this table is referred to as the {\it standard data combination} in this article. 
\begin{table}
    \centering
    \begin{tabular}{|c|c|c|}
    \hline
         Name& Likelihood &Multipole/Redshift \\    \hline    \hline
         P2018& {\texttt{commander}}& $\ell=2-29$ (TT) \\ 
         &  \texttt{SimAll} & $\ell=2-29$ (EE) \\ 
         &  \texttt{plik} TTTEEE & $\ell=30-2508$ (TTTEE)* \\
         &  lensing & $\ell=8-400$ \\ \hline
         
        P2018SR2& P2018  & Same as Planck 2018 \\ 
       &  but \texttt{SimAll} replaced&  \\

      &  by SROLL2 &  \\\hline 
 
       P2018WLFI& Pixel based &  $\ell=2-29$ (T,E) \\ 
       &  \texttt{plik} TTTEEE & $\ell=30-2508$ (TTTEE)* \\
       &  lensing & $\ell=8-400$ \\ \hline
 %

        UV17 &  &  $z=6.5-11$ \\\hline 
  UV15 &  &  $z=6.5-11$\\ \hline 
         QHII& & $z=6-8$ \\\hline
         
    \end{tabular}
    \caption{Likelihoods and multipole ranges used in this article. Hereafter, different combinations of likelihoods will be referred with the names mentioned in the first column. Second column contains the appropriate names of the likelihood, wherever applicable.*Note that for TE and EE in Plik the range is $\ell=30-1996$\\ }
    \label{tab:Datasets}
\end{table}

\begin{table}
    \centering
    \begin{tabular}{|c|c|}
    \hline
         Parameters&  Priors \\\hline\hline
         $z_{\rm re}$ & Fixed from $\tau$  \\\hline
         $\Delta z_{\rm re}$& 0.5-10  \\\hline
        $\alpha$&  0-8 \\\hline
        $\log_{10} \dot{\xi}_{\rm reion}$& 20-30  \\\hline
        $\beta$& 0-10  \\

         \hline
    \end{tabular}
    \caption{Priors used for the parameters related to reionization history.}
    \label{tab:priors}
\end{table}

We use CosmoChord~\cite{Handley:2015fda,Handley:2015vkr}, (modified to include our extended model of reionization) for nested sampling, to derive the posterior distributions and for model comparison. We use 2048 live points for the sampling. {\texttt{fgivenx}}~\cite{fgivenx} is used to obtain functional posterior from the samples.

For all our analysis, we jointly vary the cosmological parameters and nuisance parameters corresponding to Planck likelihoods. We sample the standard cosmological parameters: baryon density $\Omega_b h^2$, dark matter density $\Omega_c h^2$, the angular diameter distance to recombination $\Theta$ and the primordial power spectrum for scalar fluctuations through its amplitude $A_s$ and tilt $n_s$. Instead of using the priors used in Planck baseline analysis, we use the priors suggested in CosmoChord. Note that though in the nested sampling, tighter priors are used, we have tested that in all our analyses, all the baseline parameters have two tailed posterior distributions. The conservative flat priors on the reionization parameters are provided in~\autoref{tab:priors}. As we mentioned before, instead of sampling $z_{\rm re}$ we sample the optical depth $\tau$ for faster convergence. $z_{\rm re}$ is solved for a given $\tau$.

\begin{figure*}[!htb]
\includegraphics[width=0.5\textwidth]{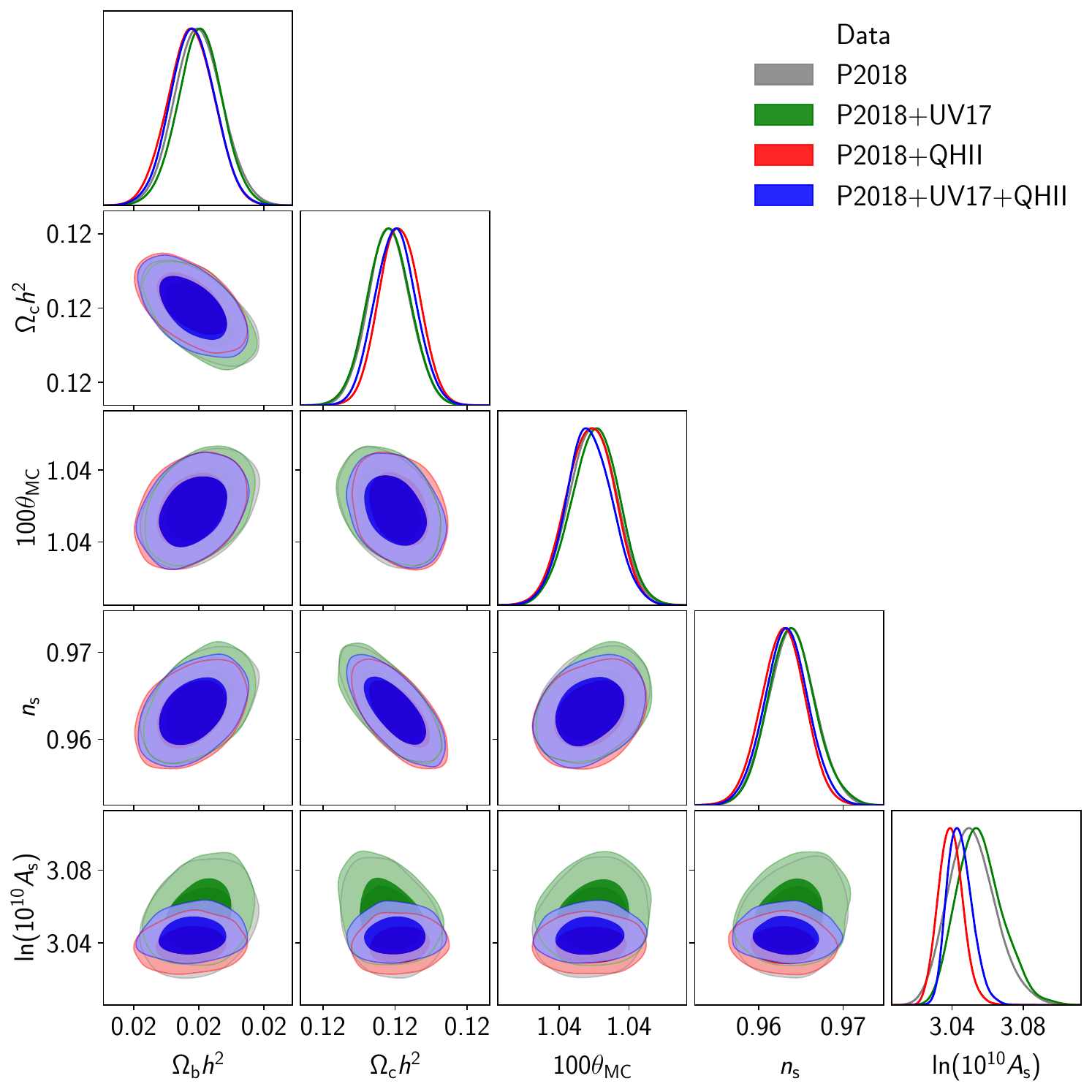}\includegraphics[width=0.5\textwidth]{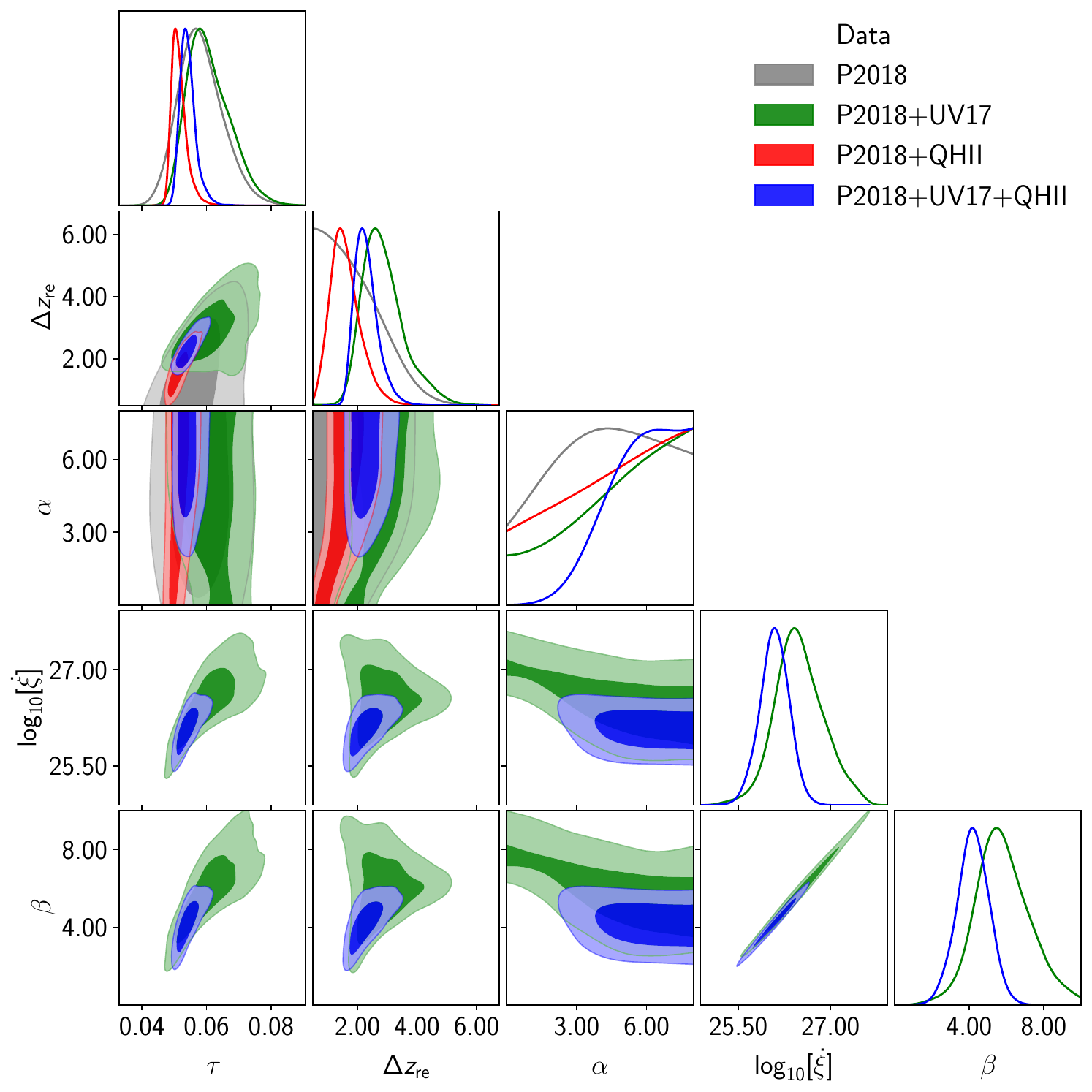}\\
\caption{Constraints on cosmological (left panel) and reionization parameters (right panel) for the standard data combinations.} \label{Fig:Asym_UV17}
\end{figure*}

\begin{figure}
\includegraphics[width=0.5\textwidth]{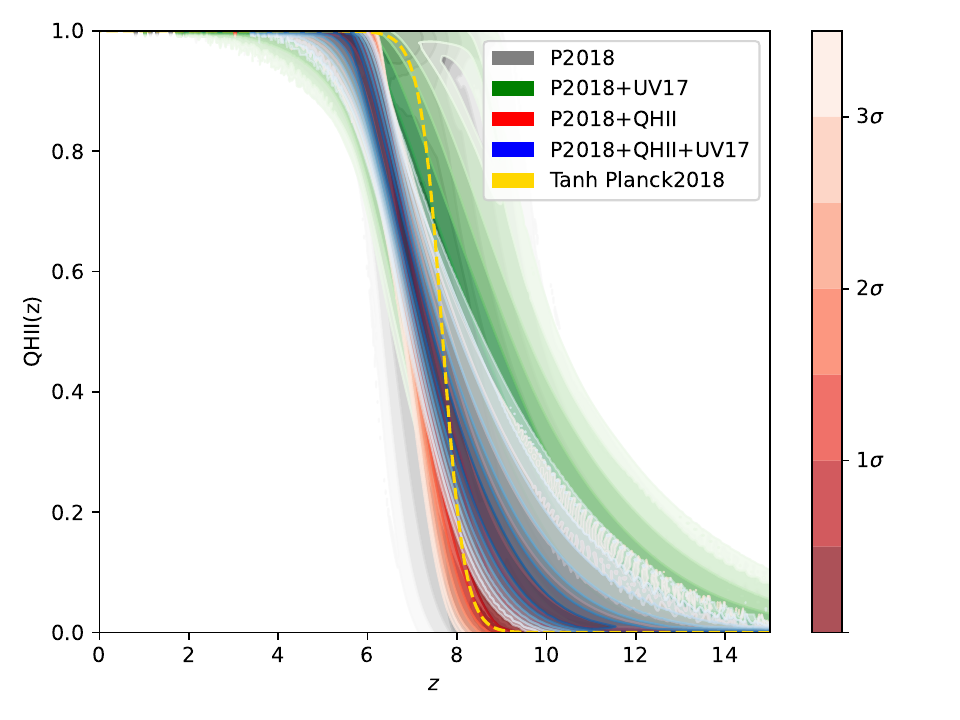}
\caption{Reconstructed reionization histories for the different dataset combinations we have considered for the baseline case of all three datasets. }\label{Fig:ReioHBaseline}
\end{figure}
\begin{figure}
\includegraphics[width=0.5\textwidth]{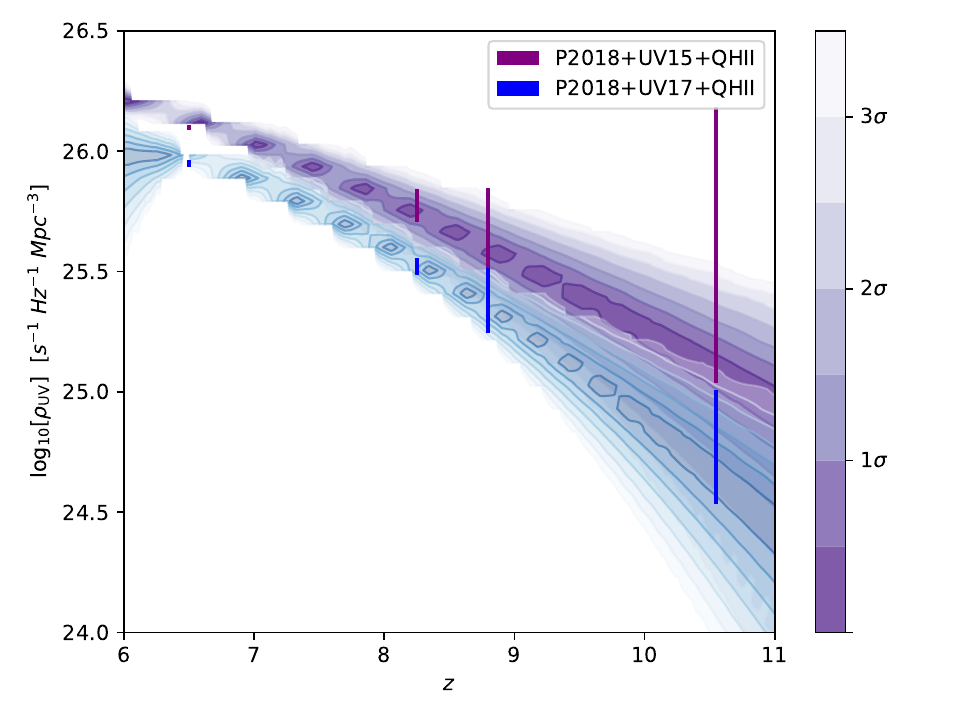}
\caption{Reconstructed UV luminosity density with the full data combinations. Comparison between the two UV luminosity density datasets UV15 in purple and UV17 in blue. Respective data points are in the same colors. }  \label{Fig:UVBaseline}
\end{figure}

In order to impose the completion of reionization by z=5 we use a soft bound on the neutral hydrogen fraction from \cite{McGreer:2014qwa} at $z=5.6$. This bound is necessary, since the variation of the reionization width allows scenarios where reionization happens till today, resulting in unrealistic bounds on the cosmological parameters.

\section{CONSTRAINTS ON EXTENDED REIONIZATION HISTORY}~\label{sec:results-I}
We derive the joint constraints on our model of reionization history and the cosmological model. We begin with the results obtained from the {\it standard dataset combinations} and then explore the constraints from alternative combinations.

\begin{table*}[!htbp]
\centering
\begin{tabular}{|c|c|c|c|c|}
\hline
 \multicolumn{5}{|c|}{Extended History of Reionization}\\
\hline
\multicolumn{1}{|c|} {Parameter} &\multicolumn{1}{|c|} {P2018} & \multicolumn{1}{|c|}  {P2018+UV17} & \multicolumn{1}{|c|} {P2018+QHII} & \multicolumn{1}{|c|} {Planck 2018+UV17+QHII} \\
\hline
$\Omega_b h^2$ &  $0.0224\pm 0.0001$ & $ 0.0224\pm 0.0001$ & $0.0224\pm 0.0001$ & $0.0224\pm 0.0001$\\
$\Omega_c h^2$ &  $0.120\pm 0.001$ & $0.120\pm 0.001$ & $0.120\pm 0.001$ & $0.120\pm 0.001$
\\
$\Theta$ &  $1.0409\pm 0.0003$ & $1.0409\pm 0.0003$ & $1.0409\pm 0.0003$ & $1.0409\pm 0.0003$
\\
$n_s$ &$0.9659\pm 0.0041$  & $ 0.9659_{-0.0043}^{+0.0039}$ & $0.9643_{-0.0039}^{+0.0038}$ & $0.9649_{-0.0039}^{+0.0038}$\\
$A_s$ &  $3.051_{-0.015}^{+0.012}$ & $ 3.056_{-0.015}^{+0.011}$& $3.039_{-0.007}^{+0.006}$ & $3.044_{-0.008}^{+0.006} $\\
$\tau$  &  $0.0582_{-0.0077}^{+0.0058}$ & $ 0.0608_{-0.0080}^{+0.0050}$ & $0.0515_{-0.0028}^{+0.0014}$ & $0.0542_{-0.0028}^{+0.0017} $
\\
${\Delta}z_{\rm re}$ &  $<2.28$ &  $ 2.88_{-0.83}^{+0.47}$ & $1.60_{-0.59}^{+0.37}$  & $2.29_{-0.46}^{+0.27} $\\
$\alpha$ &  $4.28_{-1.74}^{+3.15} (-)$ & $>3.95 (-)$& $>3.39 (-)$ & $>5.08 (>3.12)$\\
$\log_{10}[\dot{\xi}]$&  $-$ &   $ 26.49_{-0.43}^{+0.35}$ & $-$ & $26.09_{-0.22}^{+0.25} $\\
$\beta$ &  $-$ &  $ 5.80_{-1.57}^{+1.27}$& $-$ & $4.21_{-0.85}^{+0.87} $\\
\hline
$ z_{\rm re} $ &  $6.60_{-0.93}^{+0.67}$ & $ 6.04_{-0.75}^{+0.29}$  & $6.16_{-0.32}^{+0.20}$ & $5.82_{-0.19}^{+0.20} $\\
\hline
\end{tabular}
\caption{Constraints on cosmological and reionization parameters. Both errors and upper and lower bounds are at 68\% C.L. The numbers in the parentheses denote 95\% bound. Note that $z_{\rm re}$ is a parameter derived from the other parameters.}
\label{Tab:Asym_UV17}
\end{table*}

\begin{table}[!htb]
\centering
\begin{tabular}{|c|c|c|}
\hline
 \multicolumn{3}{|c|}{Extended History of Reionization}\\
\hline
\multicolumn{1}{|c|} {Parameter} &\multicolumn{1}{|c|} {P2018+UV15} & \multicolumn{1}{|c|} {P2018+UV15+QHII} \\
\hline
$\Omega_b h^2$ &  $0.0224\pm 0.0001$ & $ 0.0224\pm 0.0001$\\
$\Omega_c h^2$ & $0.119\pm 0.001$ & $ 0.120\pm 0.001$ \\
$\Theta$ &  $1.0410\pm 0.0003$ & $ 1.0409\pm 0.0003$ \\
$n_s$ & $0.9672\pm 0.0042$ & $ 0.9657\pm 0.0038$ \\
$A_s$ &   $3.067_{-0.013}^{+0.011}$ & $ 3.051_{-0.0088}^{+0.0075}$\\
$\tau$  &   $0.0667_{-0.0067}^{+0.0052}$ & $ 0.0583_{-0.0040}^{+0.0027}$  \\
${\Delta}z_{\rm re}$ &  $3.05_{-1.30}^{+0.77}$ & $ 3.16_{-0.55}^{+0.44}$\\
$\alpha$ &   $- (-)$ & $>5.79 (>2.21)$\\
$\log_{10}[\dot{\xi}]$& $26.47_{-0.42}^{+0.43}$ & $ 26.01_{-0.25}^{+0.20}$ \\
$\beta$ &  $6.58_{-1.39}^{+1.45}$ & $ 4.81_{-0.81}^{+0.64}$ \\
\hline
$ z_{\rm re} $ &  $6.63_{-1.72}^{+1.13}$ & $ 5.50_{-0.23}^{+0.12}$ \\
\hline
\end{tabular}
\caption{Parameter errors and lower bounds are 68\% C.L. whereas in parentheses we present the lower bound at 95\% C.L..}
\label{Tab:Asym_UV15}
\end{table}
\begin{table}[!htbp]
\centering
\begin{tabular}{|c|c|c|}
\hline
\multicolumn{1}{|c|} {Parameter} & \multicolumn{1}{|c|} {P2018SR2+UV17}& \multicolumn{1}{|c|} {P2018SR2+UV15} \\
\multicolumn{1}{|c|} { } & \multicolumn{1}{|c|} {+QHII}& \multicolumn{1}{|c|} {+QHII} \\
\hline
$\Omega_b h^2$ & $0.0224\pm 0.0001$& $0.0224\pm 0.0001$ \\
$\Omega_c h^2$ & $0.120\pm 0.001$& $0.120\pm 0.001$\\
$\Theta$ & $1.0409\pm 0.0003$ & $1.0410\pm 0.0003$\\
$n_s$ & $0.9648\pm 0.0039$ & $ 0.9656\pm 0.0038$ \\
$A_s$ & $3.046_{-0.008}^{+0.007}$ & $ 3.053\pm 0.008$ \\
$\tau$  & $0.0552_{-0.0031}^{+0.0018} $& $0.0586_{-0.0039}^{+0.0026}$  \\
${\Delta}z_{\rm re}$ & $2.43_{-0.49}^{+0.30} $& $3.19_{-0.52}^{+0.42}$\\
$\alpha$ & $>5.15 (>3.07)$& $>5.77 (>2.14)$\\
$\log_{10}[\dot{\xi}]$& $26.16_{-0.21}^{+0.24}$&$26.02_{-0.24}^{+0.18}$\\ 
$\beta$ & $4.48_{-0.84}^{+0.96} $& $4.83_{-0.77}^{+0.58}$\\
\hline
$ z_{\rm re} $ & $5.80_{-0.20}^{+0.21}$ & $ 5.51_{-0.23}^{+0.12}$\\
\hline
\end{tabular}
\caption{Constraints on cosmological parameters from the combination of astrophysical datasets and Planck 2018 with SROLL2 likelihood for large scale polarizarion.}
\label{Tab:Asym_UV17SROLL2}
\end{table}

\begin{figure*}[!htb]
\includegraphics[width=0.5\textwidth]{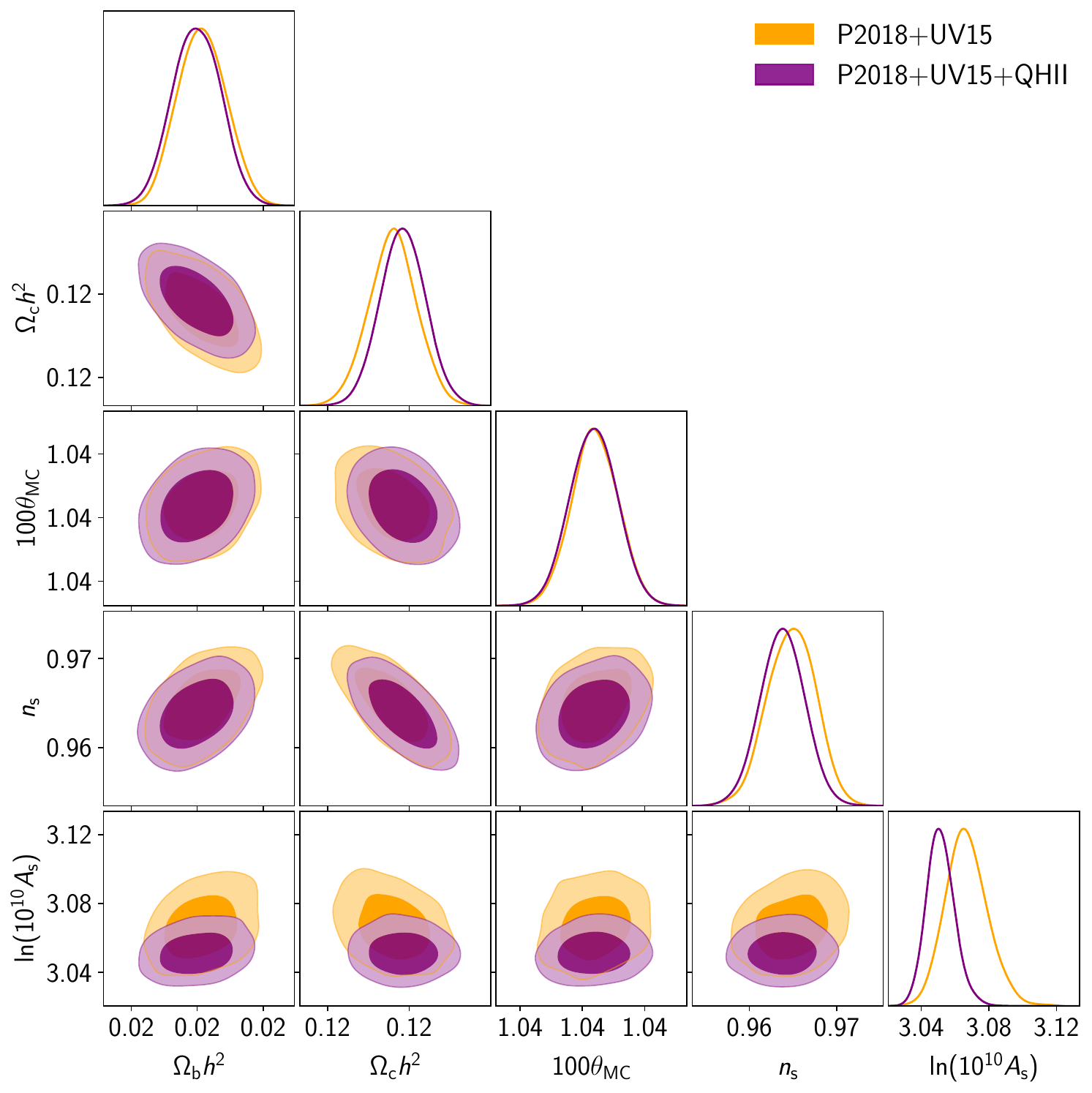}\includegraphics[width=0.5\textwidth]{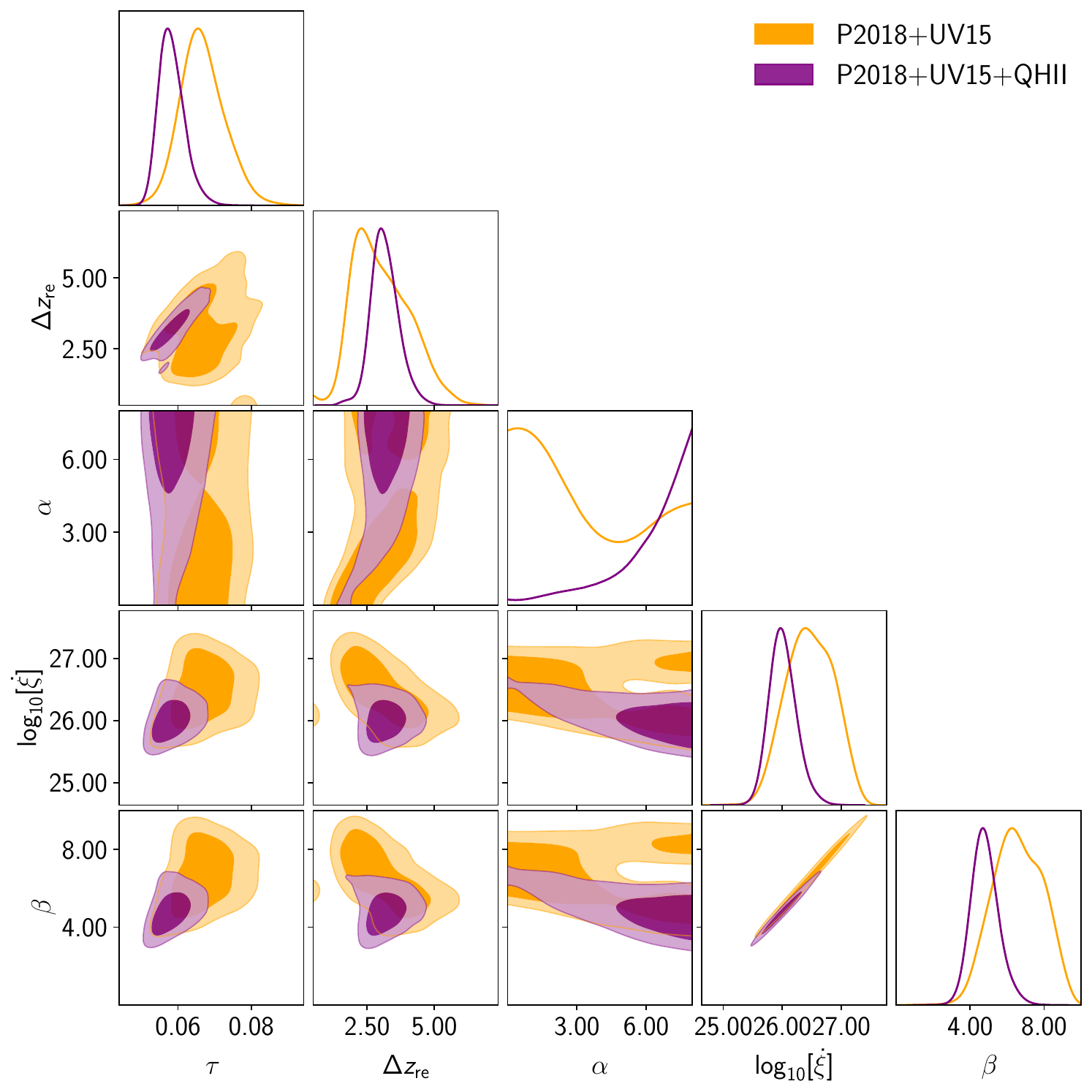}\\
\caption{Constraints on cosmological (left panel), and reionization parameters (right panel), for the data combinations considering UV15 instead of UV17. } \label{Fig:Asym_UV15}
\end{figure*}

\begin{figure}
\includegraphics[width=0.5\textwidth]{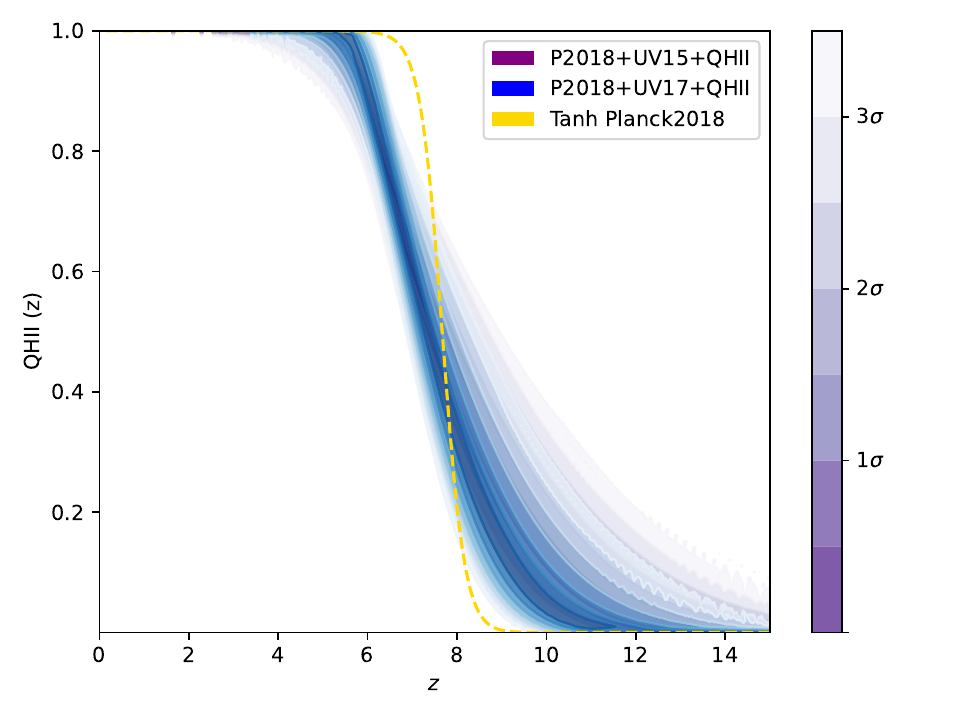}
\caption{Comparison of reconstructed reionization histories between UV with conservative cut at -17 magnitude and the more aggressive cut at -15 magnitude. } \label{Fig:ReioHBaseline2}
\end{figure}

\begin{figure*}[!htb]
\includegraphics[width=0.5\textwidth]{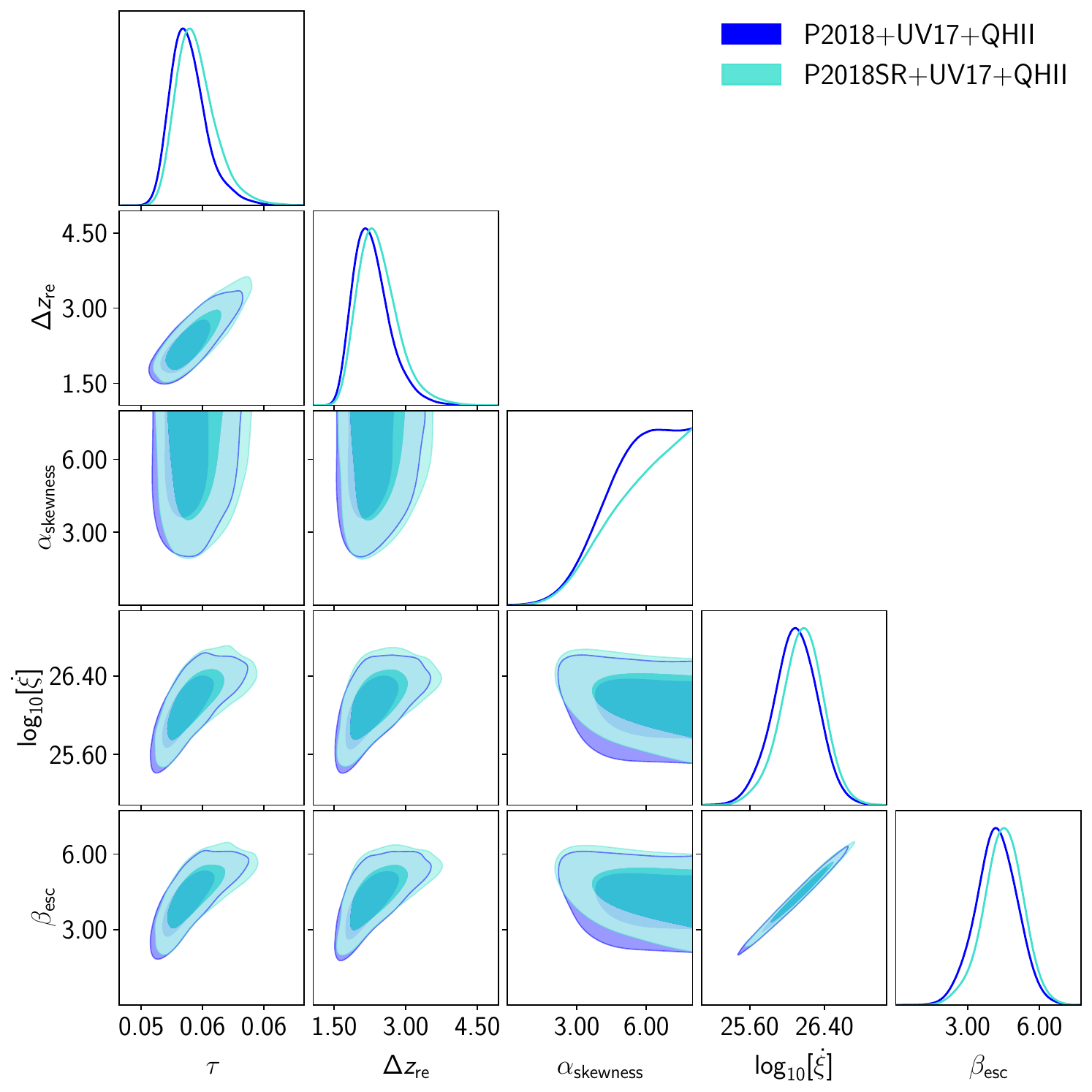}\includegraphics[width=0.5\textwidth]{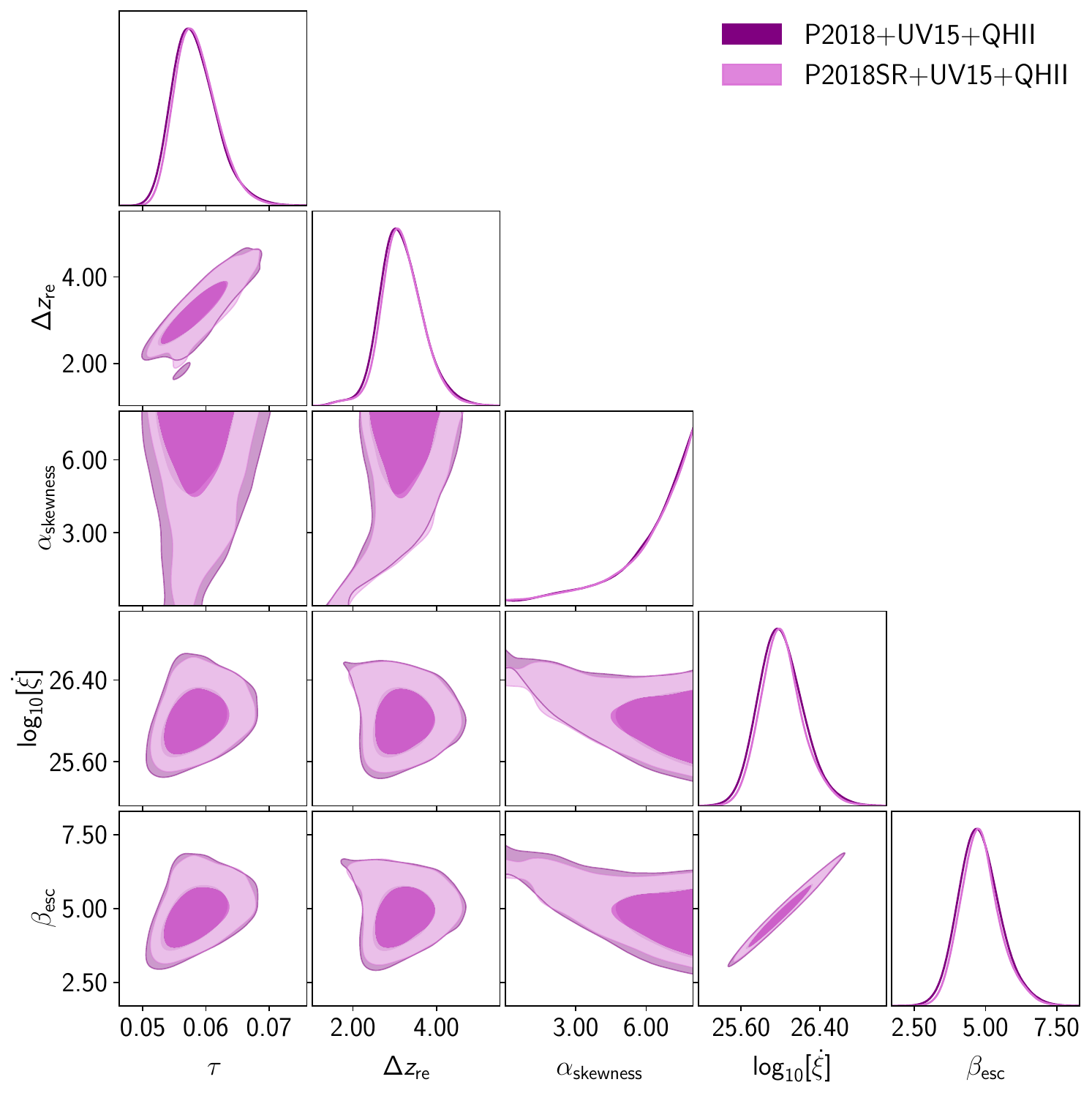}\\
\caption{Constraints on reionization parameters with Planck 2018 SROLL-2 likelihood. Left panel is the case with the combination with UV17 and QHII. The right panel is the combination considering UV15.} \label{Fig:Asym_sroll}
\end{figure*}

\begin{figure}
\includegraphics[width=0.5\textwidth]{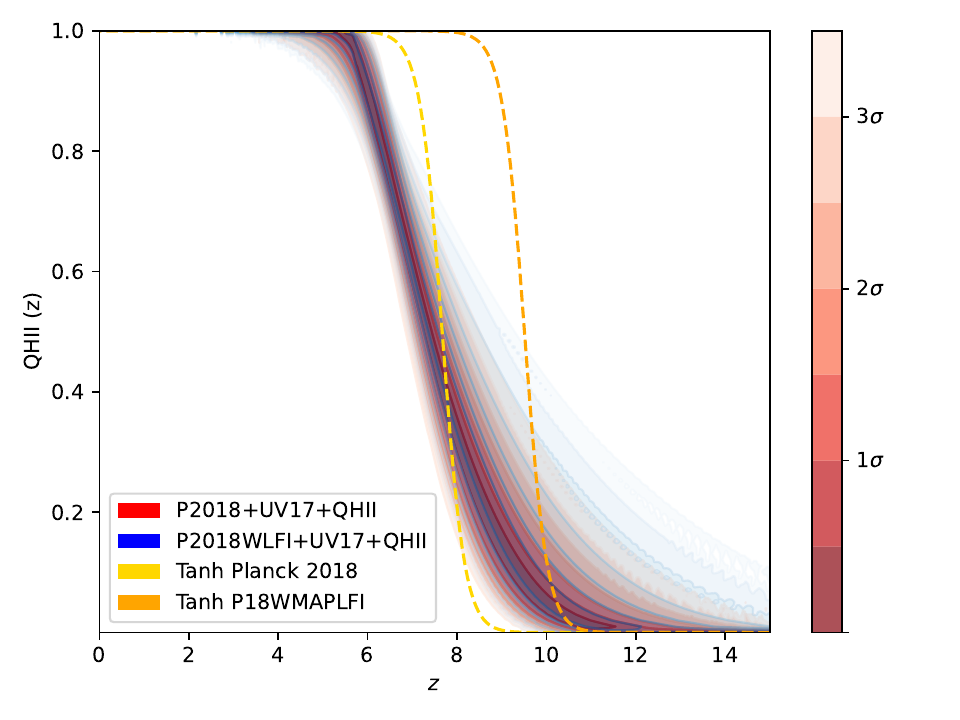}
\caption{Comparison of reconstructed reionization histories between the standard dataset combination and the same combination using the CMB large scale polarization data from WMAP+Planck LFI~\cite{Natale:WMAPLFI}, instead of Planck 2018 standard polarization.} \label{Fig:ReioHWMAPLFI}
\end{figure}

\subsection{Standard dataset combinations}
We start by investigating the different constraints from different dataset combinations. Since we vary all cosmological parameters the only common dataset to all the combinations is Planck 2018 (P2018 in tables and figures) that enables to constrain also the standard parameters together with the reionization ones. After testing the results of CMB alone we add in turn the UV luminosity density with the standard magnitude cut UV17 (in P2018+UV17), and the data from quasars and GRB (in P2018+QHII), finally we study the full combination of the three datasets (P2018+UV17+QHII).

The results are compared in~\autoref{Fig:Asym_UV17} with on the left the triangle plot for the standard cosmological parameters and on the right for the reionization history parameters.
The corresponding uncertainties and bounds are reported in~\autoref{Tab:Asym_UV17}, both are at 68\% C.L. (with 95\% C.L. in parentheses, in cases with one tailed distributions).

We find that the extended history does not change cosmological parameters from the baseline results, with the mean values being completely consistent with the Planck 2018 baseline results~\cite{Planck2018:param}. When QHII data is added, we find tighter constraint on the reionization history with a preference of shorter histories compared to P2018 and P2018+UV17. This preference results in a lowering of optical depth value with smaller error bars.

The shape of the reionization history, as expected, can not be constrained well with Planck data only. While Planck data can not distinguish between a symmetric and asymmetric models of reionization, it shows a marginal preference for an asymmetric shape as can be noted from the constraints on the parameter of asymmetry, $\alpha$. With P2018+UV17 and P2018+QHII we have lower bounds at 68\% C.L.. However, addition of these datasets put significantly tighter constraints on the duration of reionization compared to P2018 only analysis. As we mentioned before, efficiency parameter is not constrained by CMB or QHII observations unless source data UV17 is included in the combination. Constraints and bounds on all the parameters of this extended model obtained from different combinations of data are completely consistent, which allows us to combine them together. In P2018+UV17+QHII, the preference for asymmetric history from P2018+UV17 and P2018+QHII gets statististically significant where $\alpha=0$ is ruled out at more than 4$\sigma$ ($2\sigma$ lower limit of $\alpha$ being 3.12).

In~\autoref{Fig:ReioHBaseline} we show the reconstructed reionization histories for the four dataset combinations compared. We first note the difference of the preferred reionization histories with respect to standard symmetric case. The addition of UV17 to P2018 still leaves room for the symmetric model of reionization and we note that a very late completion of reionization is also allowed, given the variation of duration of reionization and soft prior on the neutral fraction. The tightening of the constrained reionization band comes from the addition of QHII data. In fact, in P2018+UV17+QHII we observe a strong rejection of a symmetric transition of the ionization fraction. We highlight that the significance of this rejection is obtained after the marginalization on $\Delta z_{\rm re}$ (along with other cosmological parameters). Therefore the result can be considered as pointing towards a strong disfavour of symmetric shape rather than limited to a particular model with fixed duration of ionization.

 The reverse engineered luminosity density is plotted in~\autoref{Fig:UVBaseline} with  blue contours corresponding to our standard dataset combination P2018+UV17+QHII compared with the data points always in blue. The remarkable consistency with the data strongly supports the concept of the reconstruction of UV densities which enables to include the constraints on the reionization sources. As in our previous works, $z_\mathrm{re}$ in \autoref{Tab:Asym_UV17} is lower than in \citep{Planck2018:param}.

\subsection{Alternative cut in the UV magnitude}
We now investigate the impact of the faint tail of the UV luminosity function by substituting UV17 dataset with the one which considers an aggressive magnitude cut at -15. Here we consider a low luminosity high redshift population contributing to the reionization history. This in our previous work shown a preference for higher optical depths and longer duration of reionization with respect to the UV17 case, implying a earlier start of ionization from the contribution of these faint sources.

The resulting posterior distributions are shown in~\autoref{Fig:Asym_UV15}. 
The limits on the parameters are tabulated in~\autoref{Tab:Asym_UV15}.
The addition of UV15 brings back the symmetrical model, this is reflected also by the posterior distribution of histories. In~\autoref{Fig:UVBaseline} posterior distribution of reconstructed UV luminosity density is plotted in purple that can be compared with the densities constrained by UV17. Apart from the considerable larger uncertainties in UV15, we also notice a shallower decrease of luminosities with redshift compared to UV17. Both these factors contribute in the increased uncertainties in $\alpha$ that makes P2018+UV15+QHII less sensitive to the shape of ionization history. Compared to UV17, here too we find larger optical depth and longer duration of reionization with larger error bars.  

\subsection{Alternative Planck large scale polarization}

Since an important contribution of reionization to CMB is at large angular scale polarization, in our analysis we also include two alternatives to the baseline low-$\ell$ E-mode likelihood of the Planck 2018 data release. The first alternative we test updates the map making algorithm to SROLL2 reducing the systematic effects at the map level~\citep{Pagano:2019tci}, for the data collected from the High Frequency Instrument (HFI). The use of this data in the Planck 2018 baseline contraints on cosmological parameters marginally increases the optical depth {\it w.r.t.} the Planck 2018 official release. The second alternative we consider is instead based on the analysis that combines WMAP large scale polarization and Planck Low Frequency Instrument polarization data as presented in~\cite{Natale:WMAPLFI}. In this case the optical depth obtained for baseline cosmology is $\tau=0.0714^{+0.0087}_{-0.0096}$.

\paragraph{Considering HFI SROLL2 polarization:} In~\autoref{Tab:Asym_UV17SROLL2} we show the constraints on cosmological and reionization parameters when we consider this alternative CMB large scale E-mode polarization likelihood. Both data combinations, P2018SR2+UV17+QHII and P2018SR2+UV15+QHII, are shown. Where UV17 data is used in the combination, we find $\alpha>5.15$ and with symmetric model ruled out at more than $3\sigma$. Following what happens for the baseline cosmology, in our extended model too, we find a marginal increase in the optical depth with $0.0551_{-0.0031}^{+0.0018}$. P2018SR2+UV15+QHII data can also rule out symmetry with $\alpha>5.77$ at 68\% C.L. and $\alpha>2.14$ at 95\%. Also for the alternative Planck baseline when we include UV15 the optical depth increases with respect to the UV17 case, P2018SR2+UV17+QHII, $\tau=0.0586_{-0.0039}^{+0.0026}$ due to its shallower decrease in the source function, allowing for earlier ionization.

\begin{table}[!htbp]
\centering
\begin{tabular}{|c|c|}
\hline
\multicolumn{1}{|c|} {Parameters}& \multicolumn{1}{|c|} {P2018WLFI+UV17} \\
\multicolumn{1}{|c|} {} & \multicolumn{1}{|c|} {+QHII}\\
\hline
$\Omega_b h^2$ & $0.0224\pm 0.0001$ \\
$\Omega_c h^2$ & $0.120\pm 0.001$\\
$\Theta$ &  $1.0409\pm 0.0003$\\
$n_s$ &  $0.9650\pm 0.0039$ \\
$A_s$ & $3.048_{-0.009}^{+0.007}$ \\
$\tau$  &  $0.0562_{-0.0040}^{+0.0019}$  \\
${\Delta}z_{\rm re}$ & $2.54_{-0.64}^{+0.29}$\\
$\alpha$ &  $>4.92 (>2.75)$\\
$\log_{10}[\dot{\xi}]$& $26.21_{-0.24}^{+0.26}$\\ 
$\beta$ &  $4.69_{-0.93}^{+0.90}$\\
\hline
$ z_{\rm re} $ &  $5.81_{-0.21}^{+0.25}$\\
\hline
\end{tabular}
\caption{Constraints on cosmological parameters from the combination of astrophysical datasets and Planck 2018 with WMAP+LFI likelihood for large scale polarizarion.}
\label{Tab:Asym_WLFI}
\end{table}

\paragraph{Considering WMAP + Planck LFI polarization:} 

This alternative likelihood at low multipoles differ from the two previously used either for the data injected and for the use of a joint TQU likelihood approach with temperature and polarization maps at the same resolution which takes into account also their cross-correlation  \cite{Finelli:2012wu,Planck2015:like}. WMAP and LFI polarization data from Planck prefers higher values of the optical depth compared to Planck HFI polarization, albeit with larger uncertainties \citep{Natale:WMAPLFI} (see also \cite{Paradiso:2022fky}). Similar to standard dataset combinations we obtain and present constraints from the combinations of P2018WLFI, UV17 and QHII in~\autoref{Tab:Asym_WLFI}. In the CMB only case (which includes also the CMB lensing) for P2018WLFI, our extended model prefers a higher optical depth $\tau=0.074^{+0.009}_{-0.010}$ compared to the hyperbolic tangent model with $\tau=0.070 \pm 0.010$. Our extended model prefers a higher optical depth compared to the hyperbolic tangent model with $\tau=0.0562^{+0.0019}_{-0.0040}$ also in the P2018WLFI+UV17+QHII analysis. Asymmetry parameter, non-zero $\alpha$ is preferred at 68\% C.L. by P2018WLFI+UV17+QHII that rejects the symmetric model at more than 3$\sigma$. A comparison of reionization history between P2018WLFI+UV17+QHII and P2018+UV17+QHII is shown in~\autoref{Fig:ReioHWMAPLFI}. Up to ($z\sim7$), we find both histories to agree. At higher redshifts, P2018WLFI allows an extended tail of reionization history compared to Planck HFI large scale polarization. This result also agrees with~\cite{PRL} where Planck HFI data strongly restricts early reionization scenarios.

\section{MODEL COMPARISON WITH THE CMB BASELINE MODEL}~\label{sec:results-II}
Our extended model embeds a symmetric model of reionization as a limit of $\alpha=0$. However, the symmetric model is not exact mathematical equivalent of hyperbolic tangent (Tanh model in \autoref{eq:Tanh}). Therefore, while from the $\alpha$, posterior, we can understand the significance of a symmetric model, we can not exactly compare with the Tanh model of reionization. To compare the Tanh model with the extended model that allows asymmetric histories, we use the Bayes' factor, which is the logarithm of the ratios of the Bayesian evidences obtained for different models for the same data combinations and same prior ranges for the common parameters. Note that for Tanh case also, we allow the duration of reionization to vary and therefore our extended model has only one extra parameter $\alpha$ compared to Tanh, that determines the asymmetry.

In~\autoref{Tab:evidence} we compare different evidences for the 4 combinations we considered in our standard dataset. In P2018 analysis and in analyses where UV17 and QHII data are used with P2018 separately, we find marginal improvement in evidence when extended model is considered compared to the Tanh model. This is expected as these dataset combinations, though preferring asymmetric histories, can not provide a strong significance (more than 95\% C.L.) for the rejection of a symmetric model. However, when P2018+UV17+QHII combination is used we find decisive evidence for asymmetric history (with $\ln [{\rm Bayes' ~factor}]]\sim 8$).

\begin{table}[!htb]
\centering
\begin{tabular}{|c|c|c|c|}
\hline
 \multicolumn{4}{|c|}{Evidences - Log(Z)}\\
\hline
 \multicolumn{4}{|c|}{Hyperbolic Tangent (\autoref{eq:Tanh})}\\
\hline
\multicolumn{1}{|c|} {P2018} & 
\multicolumn{1}{|c|} {P2018+UV17} & \multicolumn{1}{|c|} {P2018+QHII} & \multicolumn{1}{|c|} {P2018+UV17+QHII} \\
\hline
$-1435.5$  &  $-1444.6$ & $-1438.5$& $-1453.0$ \\
\hline
 \multicolumn{4}{|c|}{Extended model (\autoref{eq:extended})}\\
\hline
$-1434.6$  & $-1443.8$& $-1437.8$ & $-1446.8$ \\
\hline
\end{tabular}
\caption{Evidences (in terms of $\log(Z)$) for the different data combinations compared between the two different models of reionization history. The top panel contains the results for the hyperbolic tangent model (\autoref{eq:Tanh}) and the bottom panel represents the extended model (\autoref{eq:extended}). With one extra degrees of freedom, the extended model is marginally preferred by P2018, P2018+UV17 and P2018+QHII. In the P2018+UV17+QHI joint analysis, we note instead a strong rejection of the hyperbolic tangent model.}\label{Tab:evidence}
\end{table}

\section{Summary}~\label{sec:summary}

Previous studies have demonstrated the power of astrophysical measurements of reionization, complementary to CMB polarization, in constraining the average optical depth and the hints of reionization histories far from the Tanh assumption which is commonly adopted in Einstein-Boltzmann codes. Motivated by these studies, we have investigated how current data such as CMB anisotropy, UV luminosity from Hubble Frontier Field and neutral hydrogen from distant sources can constrain reionization beyond the average optical depth.

We have introduced a novel extended model of reionization 
to quantify the possible asymmetry of the evolution of average ionized hydrogen fraction as usually assumed in CMB studies. The ionization fraction in our model is expressed through a skewed normal distribution where the skewness parameter describes the asymmetry of the evolution of the neutral/ionized fraction as a function of redshift. Differently from our previous free-form reconstructions of the ionization history \citep{PRL,PRD}, a symmetric redshift evolution is nested in this extended model. We have reversed engineered the source function, the UV luminosity density from the extended model, that enables us to combine the source data from Hubble Frontiers Field with CMB and neutral/ionization fraction data from AGNs and GRBs even when modelling the ionization fraction evolution with redshift: note that in this way we obtain inferred confidence level contours on $x_e$ larger than in the reconstruction approach used in \cite{PRL}.
Below we highlight the main findings of our analysis:

\begin{enumerate}

\item Planck 2018 data, UV luminosity data between $z=6.5-11$ (with a limiting magnitude of -17) and neutral fraction data jointly lead to $\tau=0.0542^{+0.0017}_{-0.0028}$ at 68 \% CL and disfavour a symmetric model of reionization at more than $4\sigma$. Individually CMB and separate combinations of CMB and the other two datasets show marginal preferences for the asymmetric model, which instead is preferred only in a combined analysis.

\item Alternative Planck polarization data sets at low multipoles confirm our findings. SROLL2 \citep{Delouis:2019bub} only marginally changes the results obtained from Planck CMB 2018 polarization likelihood: the optical depth is constrained at a comparatively higher value $\tau=0.0552^{+0.0031}_{-0.0018}$ with marginal increase in the uncertainties. We have then included for the first time the WMAP and Planck LFI likelihood \citep{Natale:WMAPLFI} in our approach. When the WMAP and Planck LFI polarization data combination is used instead of Planck HFI polarization, we find a small shift in the estimated values of optical depth and reionization duration with $\tau=0.0562^{+0.0019}_{-0.0040}$, at 68 \% CL, in the joint analysis with astrophysical measurements. Both these alternate analyses too rule out a symmetric model.

\item As in \citep{PRD}, we also tested an alternative magnitude cut on the UV luminosity density, an aggressive cut at -15 that includes fainter sources. UV15 data with larger error bars and shallower decrease with redshift (compared to UV17) is less constraining and the symmetric model is disfavoured only at 95\% C.L.

\item Our extended model, allowing both symmetric and asymmetric histories, does not exactly reproduce the hyperbolic tangent model. Therefore, we perform a model comparison with Bayesian evidence between the extended model and the Tanh model with an adjustable width. Apart from the analysis with full data combination, we find that the extended model is marginally or moderately preferred by the data. However, the joint analysis with the full data combination decisively selects the extended model preferring asymmetry over the Tanh symmetric model.

\end{enumerate}

It will be interesting to test these findings with the available and upcoming data from the James Webb Space Telescope and Euclid deep field \citep{EUCLID:2011zbd,Euclid:2022bqs,2023AJ....165...13W}. Work in this direction is in progress.
\begin{acknowledgments}

DP and FF acknowledge financial support the agreement n. 2020-9-HH.0 ASI-UniRM2. 
DP, DKH and FF acknowledge the travel support through the India-Italy mobility program 'RELIC' (INT/Italy/P-39/2022 (ER)). DKH would like to acknowledge the support from CEFIPRA grant no. 6704-4. DP acknowledges the computing centre  of Cineca and INAF, under the coordination of the ``Accordo Quadro MoU per lo svolgimento di attività congiunta di ricerca Nuove frontiere in Astrofisica: HPC e Data Exploration di nuova generazione", for the availability of computing resources and support with the project INA23\_C9A05. 
The authors acknowledge  the  usage  of  computational  resources at the Institute of Mathematical Science’s High Performance Computing facility (hpc.imsc.res.in) [Kamet and Nandadevi].
GFS acknowledges Laboratoire APC-PCCP, 
and also the financial support of the UnivEarthS Labex program at Universite de Paris  (ANR-10-LABX-0023 and ANR-11-IDEX-0005-02). 
\end{acknowledgments}
\appendix
\section{}\label{app}
\begin{figure}[!htb]
\includegraphics[width=0.5\textwidth]{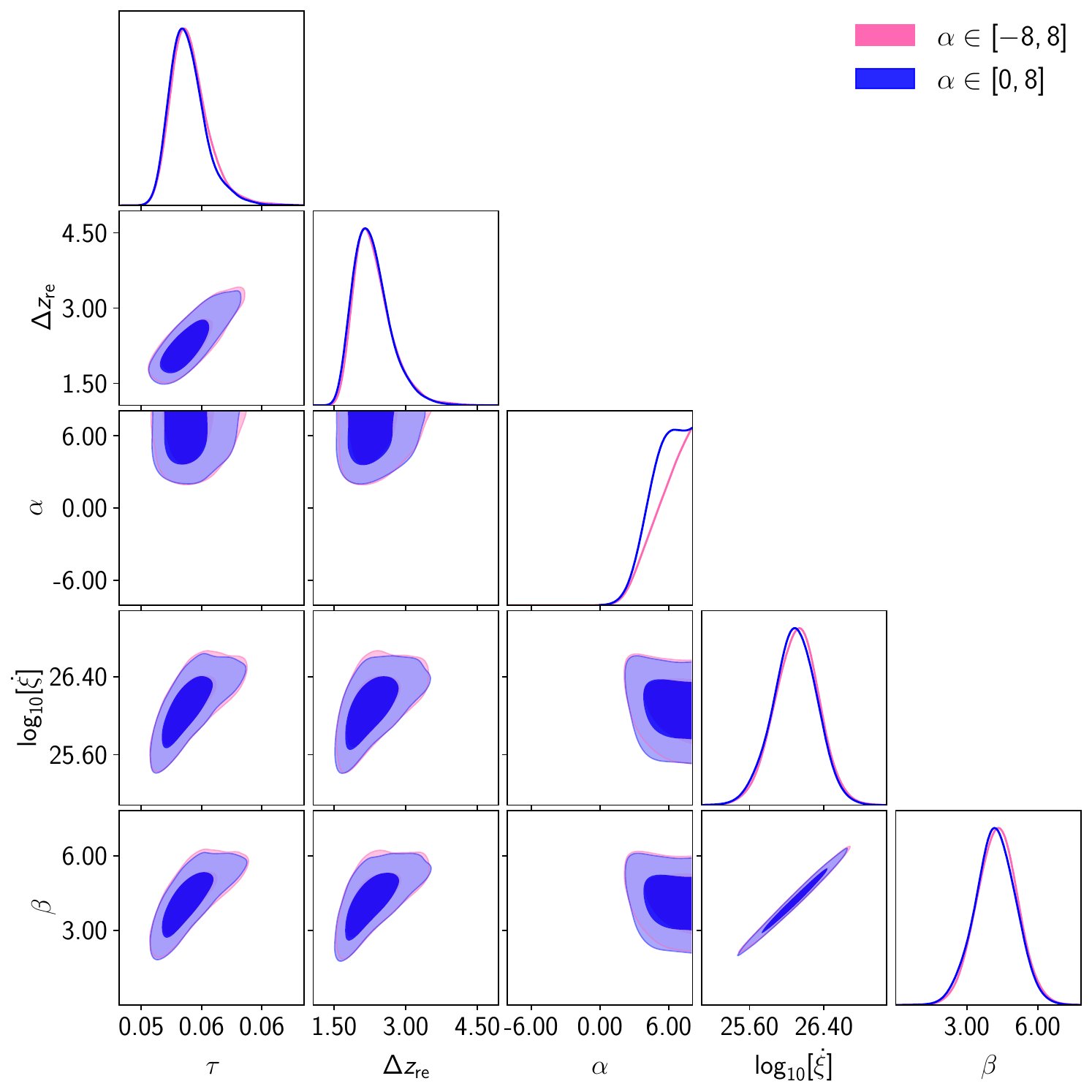}
\caption{Constraints on reionization parameters for the baseline data combination P18+UV17+QHII considering also negative values for the asymmetry parameter.} \label{Fig:Asym_m1010}
\end{figure}

The sign of the asymmetry parameter impacts the slopes at beginning and end of reionization, the negative branch has a sharper bend at the beginning of reionization whereas the positive branch has a smoother bend, the opposite happens at the end of reionization with negative branch having a shallow slope whereas the positive tends to be more abrupt.
In this appendix we test the impact of opening our prior to the negative branch for the asymmetry parameter $\alpha \in$ [$-$8,8].
This gives for the standard data combination again a strong preference for asymmetric models with positive $\alpha$, with $\alpha>5.28, >3.07$ at 68 and 95 \% C.L. respectively. In \autoref{Fig:Asym_m1010} we show the two-dimensional posteriors for the reionization parameters compared with the positive prior case.

\bibliography{reionref}

\end{document}